\documentclass[aps,prd,groupedaddress]{revtex4}
\usepackage{amssymb}
\usepackage{amsmath}
\usepackage{graphicx}
\usepackage{epstopdf}
\usepackage{xcolor}

\newcommand{\half} {{\textstyle \frac{1}{2}}}

\begin{document}

\title{Finite-Distance Gravitational Lensing of a Global Monopole in a Schwarzschild--de Sitter Spacetime}

\author{Yi Lu}
\author{Xiao-Yin Pan}
\email{panxiaoyin@nbu.edu.cn}
\affiliation{Department of Physics, Ningbo University, Ningbo 315211, Zhejiang Province, China}

\author{Meng-Yun Lai}
\email{mengyunlai@jxnu.edu.cn}
\affiliation{College of Physics and Communication Electronics, Jiangxi Normal University, Nanchang 330022,  Jiangxi Province, China}

\author{Qing-hai Wang}
\email{qhwang@nus.edu.sg}
\affiliation{Department of Physics, National University of Singapore, Singapore 117551,  Singapore}

\date{\today}

\begin{abstract}
We investigate the gravitational lensing of a Schwarzschild--de Sitter black hole with a global monopole at finite distances. In this asymptotically nonflat spacetime, the deflection angle of light is decomposed into two parts: the first derives from the orbit differential equation, and the second originates from the metric itself. By absorbing the cosmological constant into the effective impact parameter, we derive an analytical expression for the first part using elliptic integrals. Combined with the second part, we obtain a complete exact solution for the deflection angle in this context. Additionally, considering that the distances from the source to the observer are large, we derive expressions for the light deflection angle in both the weak- and the strong-field limits. In both cases, we find that the deflection is enhanced by the presence of the global monopole, further supporting its potential role as an alternative to elusive dark matter.
\end{abstract}


\maketitle

\section{Introduction}
\label{sec:intro}

The deflection of light is one of the earliest predictions and played a significant role in the initial verification of Einstein's theory of general relativity \cite{1,2}. Building on the concept of light deflection, gravitational lensing has become an indispensable tool in astrophysics and cosmology \cite{Ovgun2019PRD,Ovgun2022JCAP,Ovgun2022EPJC,Virbhadra2024PRD,Virbhadra2022GRG,Virbhadra2024CanJPhys,Virbhadra2022PRD,Ovgun2025PLB}. It is utilized for measuring the mass of galaxies or galaxy clusters, determining the Hubble constant, and studying the properties of dark objects and dark energy \cite{3,4,Virbhadra2022GRG,Virbhadra2024CanJPhys,Ovgun2025PLB}. In particular, dark energy is widely believed to be responsible for the accelerated expansion of our Universe, as discussed in Refs.~\cite{5,6,7,8} and the references therein. To understand the nature of dark energy, researchers are interested in examining how different candidates vary in their observable effects. Although its nature remains inconclusive, the cosmological constant $\Lambda$ has been a leading candidate, with various authors exploring its potential contribution to the bending of light or gravitational lensing.

There has been an ongoing debate regarding whether the cosmological constant plays a role in gravitational lensing. Computations in Schwarzschild--de Sitter (SdS) or Kottler spacetime \cite{9} reveal that $\Lambda$ does not appear in the second-order differential equation of the photon (null geodesics). Various authors, as seen in Refs.~\cite{10,11,12,13,14}, have argued that $\Lambda$ does not contribute to lensing. However, Rindler and Ishak, utilizing the SdS metric and the invariant formula of cosine, pointed out that $\Lambda$ does indeed affect the bending of light \cite{15}. They argued that the differential equation and its integral represent only half of the story, with the other half being the metric itself. This perspective was supported by many subsequent studies in various ways and levels of generality \cite{16,17,18,19,20,22,23,Guenouche1,Guenouche2,Guenouche3}. Among these, Arakida and Kasai were the first to realize that the contribution of the cosmological constant can be absorbed into the definition of the impact parameter, concluding that the bending angle of light does not change its form even if $\Lambda \neq 0$ \cite{23}.

Ishihara \textit{et al.}~highlighted that previous works employing the deflection angle of light in an asymptotically flat spacetime commonly referred to as Weinberg's method--—are insufficient for addressing asymptotically nonflat spacetimes when the cosmological constant is present \cite{24,24a}. Applying the Gauss-Bonnet theorem with the optical metric, they redefined the deflection angle of light in an asymptotically nonflat spacetime by assuming a finite distance from the source to the observer. This new formulation was further enhanced from the observer's perspective by Takizawa \textit{et al.}~\cite{25}. In this approach, the deflection angle is divided into two parts: the first arises from the photon's differential equation and maintains the same integral form as in an asymptotically flat spacetime, while the second part is derived from the metric itself. This formulation has been applied to investigate various systems, including SdS spacetime \cite{25}, Weyl gravity \cite{26}, Reissner--Nordstr\"om--de Sitter spacetime \cite{28}, a rotating wormhole \cite{29}, and a rotating global monopole \cite{30}. For a comprehensive review on this topic, please refer to Ref.~\cite{31}.

Dark matter is a critical component of the leading standard cosmological model, $\Lambda$CDM \cite{32,33,34,35,36}. With repeated null results in earth-based experiments searching for dark matter, alternative theories are also being explored. Examples include MOdified Newtonian Dynamics (MOND) \cite{37,38}, conformal gravity \cite{39,40}, global monopoles \cite{45}, and so on. Global monopoles are topological defects in the vacuum manifold, potentially produced by phase transitions in the early universe due to the breakdown of global gauge symmetries \cite{42,43}. An intriguing property of global monopoles is that their energy density decreases with distance as $r^{-2}$ \cite{44}, a characteristic exploited to explain the flat rotation curves of stars and gases in the outer regions of several galaxies \cite{45}. In addition to other appealing properties, such as their contribution to the average density of the universe being of the right order of magnitude to account for the inflationary prediction of a universe with critical density, the concept of global monopoles as seeds for galaxy formation and models for galactic dark matter has been proposed \cite{45,46,47}. Global monopoles have also been extensively studied for various implications of their gravitational effects \cite{48,49,50,51,52,53,Gussmann,Pereira1,Pereira2,Pereira3,Pereira4}.

Following Barriola and Vilenkin's discovery of the first static spherically symmetric black hole with a global monopole \cite{44}, various physical properties of different black-hole--global-monopole systems have been extensively studied \cite{54,55,56,57,58,59,60,61,62,63,64,65,66}. In particular, strong and weak gravitational lensing has been investigated for different black-hole--global-monopole systems. For example, the strong-field gravitational lensing of a Schwarzschild black hole with a global monopole, as well as with both ordinary and phantom global monopoles, has been explored in Refs.~\cite{62} and \cite{63}, respectively. To test the alternative dark matter model based on global monopoles, the weak field gravitational lensing of a Schwarzschild black hole with a global monopole was examined in Ref.~\cite{64}. In the context of $f(R)$ gravity, the strong field gravitational lensing of a Schwarzschild black hole with a global monopole was investigated in Ref.~\cite{65}. Additionally, the light deflection angle for a Reissner-Nordstr\"{o}m-de Sitter black hole in the gravitational monopole background was studied using the Rindler-Ishak method in Ref.~\cite{66}.

In this paper, we investigate the gravitational light bending in the context of a Schwarzschild--de Sitter--black-hole--global-monopole system, with a focus on the effects that the global monopole has on the deflection angle of light in such an asymptotically nonflat spacetime. A nontrivial characteristic of this system is its solid deficit angle \cite{44}. By employing the formulation developed in Refs.~\cite{24,24a,25}, and noting that the first part of the light deflection angle retains the same form as in the absence of $\Lambda$, we compute the light deflection angle for this system. We derive an exact analytical expression for the first part of the light deflection angle using elliptic integrals, based on analytically obtainable roots of a cubic polynomial. By combining this with the second part from the metric, we determine the complete exact deflection angle of light for the system.

To compare our findings with observations, we perform a detailed analysis of both the weak-field and the strong-field limits of the derived exact deflection angle. In the weak-field approximation, we find that the presence of a global monopole enhances the gravitational lensing effect beyond standard predictions. In the strong-field regime, where light trajectories can loop around the massive object multiple times, the influence of the global monopole becomes even more significant, further amplifying the deflection of light and increasing the angular separation of relativistic images. This enhancement in gravitational lensing, attributed to the global monopole, provides compelling evidence supporting the consideration of the global monopole as a viable alternative to dark matter.

The plan of the paper as the follows. In Sec.~\ref{sec:setup}, we give a brief overview of the  Schwarzschild--de Sitter black hole  with a global monopole, and then we proceed to solve the null geodesic equation and obtain an exact analytical expression for the light deflection angle in Sec.~\ref{sec:exact}. Based on this exact result, then we study the weak- and the strong-field limits of the gravitational lensing in Secs.~\ref{sec:weak} and \ref{sec:strong}, respectively.  Section \ref{sec:observation} is devoted to explore some observables of the relativistic images, and the results are summarized in Sec.~\ref{sec:summary}. In Appendix \ref{sec:appendix}, the weak-field limit of the deflection angle of light is independently verified through a direct perturbative calculation based on the Gauss-Bonnet theorem for the case of $\Lambda = 0$. In Appendix~\ref{sec:appendixB}, we repeat this computation in Schwarzschild--de Sitter spacetime with $\Lambda \neq 0$.

Throughout the paper we use metric signature $(-,+,+,+)$ and the natural units $G=c=1$, where $G$ is  Newton's constant and $c$ is the speed of light.

\section{Schwarzschild--de Sitter Black Hole with a Global Monopole}
\label{sec:setup}

The spacetime arena for our investigation of light deflection is the Schwarzschild--de Sitter or Kottler black hole \cite{9}, set against the backdrop of a global monopole. The action of a Schwarzschild--de Sitter black hole is expressed as
\begin{align}
	S_{\text{E}}=\frac{1}{16\pi G}\int d^4x \sqrt{-g}(R-2\Lambda),
\end{align}
where $g_{\mu\nu}$ is the spacetime metric tensor, $g$ is its determinant, $R$ is the Ricci curvature scalar, and $\Lambda$ is the cosmological constant, which is very small ($\Lambda\sim 10^{-52}\, \text{m}^{-2}$, as estimated by large scale structure observations \cite{67}). The total action of the system we analyze is given by
\begin{align}
	S=S_{\text{E}} +S_{\text{GM}},
\end{align}
where $S_{\text{GM}}$ describes a global monopole,
\begin{align}
	S_{\text{GM}}=-\int d^4x \sqrt{-g} \mathcal{L}_{\text{GM}},
\end{align}
with
\begin{align}
	\mathcal{L}_{\text{GM}}=\frac{1}{2} (\partial_\mu \chi^i) (\partial^\mu \chi^i)+\frac{\lambda_0}{4}(\chi^i\chi^i-\eta^2)^2
\end{align}
being the Lagrangian density of the global monopole \cite{32}. Here $\lambda_0$ is the self-coupling term, $\eta\sim 10^{-3}$ is the scale of the gauge symmetry breaking parameter, and $\chi^i$ is a triplet scalar field with $i=1,2,3$. It is given by
\begin{align}
	\chi^i=\eta h(r)\frac{x^i}{r},
\end{align}
where $x^i$ are the spatial Cartesian coordinates, with $r=\sqrt{x^ix^i}$ representing the radial distance. Here, $h(r)\rightarrow 1$ as $r\gg \delta$ with $\delta\sim \lambda_0^{-\frac{1}{2}} \eta^{-1}$ being the order of core size of the global monopole. The effective mass of the global monopole is negative, and is very small on the astrophysical scale; hence, it is usually neglected.

By applying the action principle, we derive Einstein's field equation
\begin{align}
	G_{\mu\nu}+\Lambda g_{\mu\nu}=8\pi  T^\chi_{\mu\nu},
\end{align}
where $G_{\mu\nu}\equiv R_{\mu\nu} - \half R g_{\mu\nu}$ is the Einstein tensor, $R_{\mu\nu}$ is the Ricci curvature tensor, and
\begin{equation}
	T^\chi_{\mu\nu}=(\partial_\mu \chi^i) (\partial_\nu \chi^i)-g_{\mu\nu} \mathcal{L}_{\text{GM}},
\end{equation}
is the energy-momentum tensor for the global-monopole field.

Assuming $h(r)\approx 1$ outside the core of the global monopole, we find that the equation of motion admits a spherical symmetric Schwarzschild--de Sitter black hole solution as a special case discussed in Refs.~\cite{60,68},
\begin{align}
	ds^2=-\mathcal{A}(r)dt^2+\frac{dr^2}{\mathcal{A}(r)}+r^2 (d\theta^2+ \sin^2\theta d\varphi^2 ),
	\label{eqn:ds2}
\end{align}
where the metric function is given by
\begin{align}
	\mathcal{A}(r)=\mathcal{A}_0-\frac{\Lambda{r}^2}{3}, \quad r\gg \delta,
	\label{eqn:metricA}
\end{align}
with
\begin{equation}
	\mathcal{A}_0(r)=a^2- \frac{2  M}{r}, \quad r\gg \delta,
	\label{eqn:metricA0}
\end{equation}
representing the metric function in the absence of cosmological constant $\Lambda$. Here, $M$ is the mass of the black hole, and
\begin{align}
	a \equiv \sqrt{1-8\pi \eta^2}	
\end{align}
is the global-monopole parameter, which satisfies $0<a\leq 1$. For $M=0$, the de Sitter horizon occurs at
\begin{align}
	r_\Lambda \equiv a\sqrt{\frac{3}{\Lambda}}.
\end{align}
It is evident that this metric function, as described in Eq.~(\ref{eqn:metricA}), characterizes the far field of a global monopole embedded in a Schwarzschild--de Sitter black hole.

The event horizons of the spacetime system are determined by the condition $\mathcal{A}(r)=0$. When $\Lambda= 0$, we get one horizon at $R_{+} = {2M}/{a^2}$. When $\Lambda > 0$, we get the algebraic equation of the form:
\begin{equation}
	\Lambda r^3 - 3a^2 r + 6M =0.
\end{equation}
If the condition $\Lambda M^2 < {a^6}/{9}$ is met---which is usually satisfied since $\Lambda M^2_{\bigodot} \sim  10^{-46}$ with $M_{\bigodot}$ representing the solar mass---the above equation yields three real roots:
\begin{align}
	 R_\iota &= \frac{2\sqrt{2}}{\sqrt{\Lambda}} a \cos \left(\varpi+\frac{2\pi}{3}\iota\right), \qquad \iota=0,1,2,
\end{align}
with $$\varpi \equiv \frac{1}{3}\cos^{-1}\left(\frac{3 M \sqrt{\Lambda}}{a\sqrt{a}}\right).$$
It is straightforward to verify that $R_1, R_3>0$ and $R_2<0$, indicating that the system has two horizons. In the absence of the global monopole ($a=1$), these expressions for the horizons reduce to the results found in Ref.~\cite{69}.

On the other hand, in the absence of the black hole ($M=0$), the line element from Eq.~(\ref{eqn:ds2}) can be recast into the form
\begin{align*}
	ds^2=-dt^2+{dr^2}+a^2r^2 (d\theta^2+ \sin^2\theta d\varphi^2 ).
\end{align*}
This indicates that the spacetime is asymptotically nonflat with a conic singularity at $r=0$, characterized by a deficit solid angle resulting from the nonzero gauge symmetry breaking, quantified by $8\pi \eta^2$ \cite{44}. As discussed in Refs.~\cite{70,71}, because $\eta$ is much smaller than the Plank scale, in physically relevant cases, the solid deficit angle is very small ($\eta\sim 10^{-3}$, $8\pi\eta^2 \sim 10^{-5}\ll 1$). With this in mind, we now proceed to calculate the deflection angle of light.

\section{Calculation of the Deflection Angle of Light}
\label{sec:exact}

To calculate the bending angle, we start by deriving the equation of motion for photons (null geodesics). Since the photon motion occurs within a plane, we use spherical coordinates $(r,\theta, \varphi)$ and, without loss of generality, select the equatorial plane by setting $\theta=\frac{\pi}{2}$. Note that the energy $E$ and the angular momentum $L$ per unit mass are conserved, and we have
\begin{equation}
	\frac{E^2}{\mathcal{A}}-\frac{\dot{r}^2}{\mathcal{A}}-\frac{L^2}{ r^2}=0,  \qquad {\dot{\varphi}}=\frac{L}{r^2},
	\label{eqn:EoM}
\end{equation}
where the dots denote the derivatives with respective to the affine parameter. From the above equation we obtain the orbit equation for the photon
\begin{align}
	\dot{r}^2=E^2-\frac{L^2 }{r^2}\left(a^2 - \frac{2M}{r} -\frac{\Lambda r^2}{3}\right)\equiv V_{\text{eff}}(r).
	\label{eqn:rdot}
\end{align}

According to the orbit equation, a photon originating from infinity with an impact parameter greater than a certain minimum value, $b > b_c$, will approach the central object and then recede after reaching a minimum distance $r_0$. If $b$ is below this threshold, the photon will be captured by the black hole. The impact parameter of the light ray is defined as
\begin{equation}
	b\equiv\frac{L}{E}=\frac{r_0}{\sqrt{\mathcal{A}(r_0)}},
	\label{eqn:bLE}
\end{equation}
where $r_0$ is the turning point of the trajectory, satisfying
\begin{equation}
	0=V_{\text{eff}}(r_0) = E^2-\frac{L^2 }{r_0^2}\left(a^2 - \frac{2M}{r_0} - \frac{\Lambda r_0^2}{3}\right).
\end{equation}
This leads to
\begin{equation}
	\frac{1}{b^2}=\frac{1}{r_0^2}\left(a^2 - \frac{2M}{r_0}\right)-\frac{\Lambda }{3}.
	\label{eqn:b}
\end{equation}
As first noted in Ref.~\cite{23}, the cosmological constant $\Lambda$ can be absorbed by introducing an effective impact parameter $b_{\text{eff}}$ as follows:
\begin{equation}
	\frac{1 }{b_{\text{eff}}^2} \equiv \frac{1}{b^2}+\frac{\Lambda }{3}
	 =\frac{1}{r_0^2}\left(a^2 - \frac{2M}{r_0}\right).
	\label{eqn:beff}
\end{equation}
It is evident that the form of Eq.~(\ref{eqn:beff}) is identical to the case without the cosmological constant, except that $b$ is replaced by $b_{\text{eff}}$, which is slightly smaller than  $b$ since $\Lambda b^2$ is a very small positive number. Considering $r$ as a function of $\varphi$, we derive the geodesic equation for the equatorial plane from Eqs.~(\ref{eqn:EoM}), (\ref{eqn:rdot}), and (\ref{eqn:beff}) as
\begin{align}
	\left(\frac{dr}{d\varphi}\right)^2
	&= \frac{ r^4}{b_{\text{eff}}^2}+2Mr-a^2 r^2.
	\label{eqn:drdphi}
\end{align}

Before deriving the deflection angle, let us take a brief detour to determine the minimum value of the impact parameter, $b_c$. The turning point $r_0$ has a minimum value $r_c$, known as the photon sphere radius, which satisfies the conditions $V'_{\text{eff}}(r_c)=0$ and $V''_{\text{eff}}(r_c)>0$. Through simple algebraic manipulation, it can be shown that the photon sphere radius is given by
\begin{equation}
	r_c=\frac{3M}{a^2}.
	\label{eqn:rc}
\end{equation}
Notably, $r_c$ is independent of the cosmological constant $\Lambda$. The critical impact parameter $b_c$ is defined as
\begin{equation}
	b_c\equiv b(r_c)=\frac{r_c}{\sqrt{\mathcal{A}(r_c)}}.
\end{equation}
By combining the above equation with Eqs.~(\ref{eqn:metricA}) and (\ref{eqn:b}), we obtain
\begin{equation}
	\frac{1 }{b^2(r_c)}= \frac{1 }{b_{\text{eff}}^2(r_c)}-\frac{\Lambda }{3},
\end{equation}
where
\begin{equation}
	b_{\text{eff}}(r_c)=\frac{r_c}{\sqrt{\mathcal{A}_0(r_c)}}=\frac{3\sqrt{3}M}{a^3},
	\label{eqn:beffrc}
\end{equation}
is the critical impact parameter in the absence of the cosmological constant.

\begin{figure}[htp]
	\centering
	\includegraphics[scale=0.6]{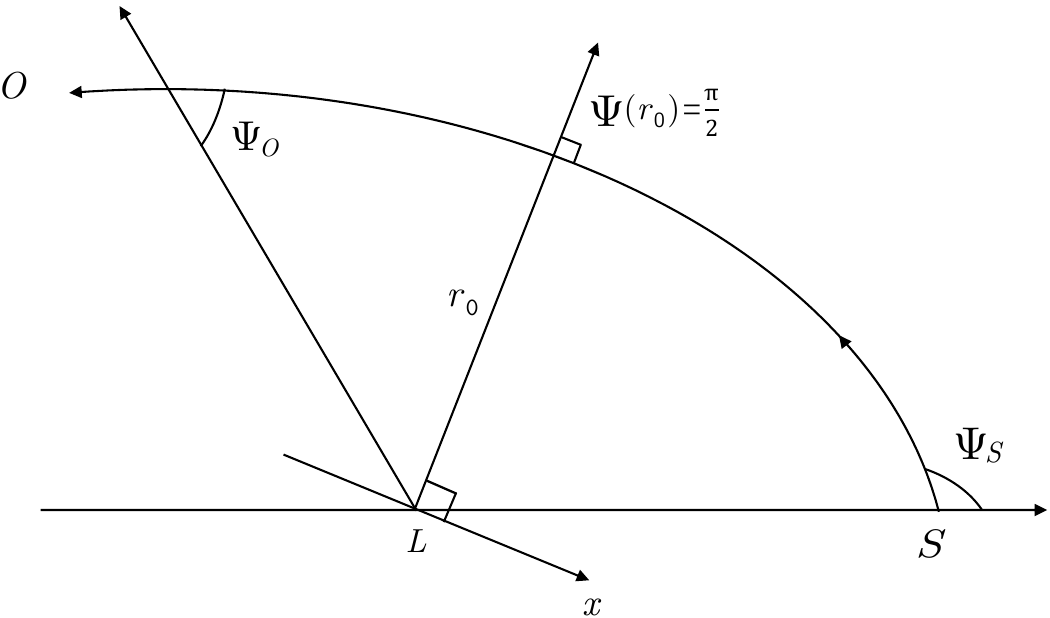}
	\caption{Lens diagram illustrating the positions of the observer ($O$), the lens ($L$), and the source ($S$). The minimum radial distance is denoted by $r_0$. The angles $\Psi_O$ and $\Psi_S$ represent the angles between the direction of light and the coordinate line at the positions of the observer and source, respectively.}
	\label{fig:lens}
\end{figure}

Returning to our main task of calculating the deflection angle, and following the methodology outlined in Refs.~\cite{24,25}, we denote the observer's position as $(r_O, \varphi_O)$ and the source's position as $(r_S, \varphi_S)$. Given that the spacetime is asymptotically nonflat, the deflection angle can be decomposed into two parts:
\begin{equation}
	\alpha_D(b,\Lambda)=\alpha_{1D}(b,\Lambda)+\alpha_{2D}(b,\Lambda).
	\label{eqn:alpha}
\end{equation}
The first part, $\alpha_{1D}(b,\Lambda)$, is derived from the orbit differential equation (\ref{eqn:drdphi}) and is expressed as
\begin{equation}
	\alpha_{1D}(b,\Lambda)=\varphi_O(b,\Lambda)-\varphi_S(b,\Lambda) \equiv \varphi_{OS}.
	\label{eqn:alpha1D}
\end{equation}
By introducing the effective impact factor, the first part of the deflection angle simplifies to a function of a single-variable $b_{\text{eff}}$, thereby eliminating the explicit dependence on the cosmological constant:
\begin{equation}
	\alpha_{1D}(b,\Lambda)=\alpha_{1D}(b_{\text{eff}}).
	\label{eqn:alpha1beff}
\end{equation}
It is important to note that the relation in Eq.~(\ref{eqn:alpha1beff}) applies only to the first part of the deflection angle. The effect of $\Lambda$ is explicitly present in the second part, which arises from the metric itself and is given by
\begin{equation}
	\alpha_{2D}(b,\Lambda)=\Psi_O(b,\Lambda)-\Psi_S(b,\Lambda).
	\label{eqn:alpha2}
\end{equation}
Here, $\Psi$ represents the angle of the light ray measured from the radial direction. Thus, $\Psi_O$ and $\Psi_S$ are the angles measured at the observer and source positions, respectively, and they can be determined as follows \cite{24}:
\begin{mathletters}
\begin{align}
\begin{split}
	\Psi_O(b,\Lambda) = \sin^{-1} \left[\frac{b}{r_O} \sqrt{ \mathcal{A}(r_O)} \right], \qquad
	\Psi_S(b,\Lambda) = \pi - \sin^{-1} \left[\frac{b}{r_S} \sqrt{ \mathcal{A}(r_S)} \right].  
\end{split}
\label{eqn:Psis}
\end{align}
\end{mathletters}
In an asymptotically flat spacetime, assuming $r_O\rightarrow \infty$ and $r_S\rightarrow \infty$, the difference $\Psi_O-\Psi_S$ approaches $-\pi$, and Eq.~(\ref{eqn:alpha}) reduces to Weinberg's expression for the deflection angle of light. Additionally, the angle $\Psi$ at the turning point $r_0$ on the light ray is given by
$$\Psi(r_0) = \sin^{-1}\left[\frac{b}{r_0}\sqrt{ \mathcal{A}(r_0)} \right]=\frac{\pi}{2}.$$
This is expected because the light ray should be perpendicular to the radial direction at the turning point, as illustrated in Fig.~\ref{fig:lens}.

With the second part of the light deflection angle given by Eqs.~(\ref{eqn:alpha2}) and (\ref{eqn:Psis}), the problem reduces to calculating the first part. To achieve this, we change the variables to $u=1/r$ and $u_0=1/r_0$, and factorize the result as follows:
\begin{align}
\begin{split}
	\left(\frac{du}{d\varphi}\right)^2 
	&= \frac{1 }{ b_{\text{eff}}^2 }-a^2 u^2+2M u^3	
	= 2M u^3-a^2 u^2+a^2 u_0^2-2M u_0^3 
	= 2M (u-u_1)(u-u_2)(u-u_3). 
\end{split}	
\label{eqn:dudphi}
\end{align}
Without loss of generality, the three roots are ranked as $u_1\le u_2\leq u_3$. These roots can be easily determined from the second line of the above equation:
\begin{align}
\begin{split}
	u_{1} &= \frac{(a^2-2M u_0)-\sqrt{a^4 + 4M a^2 u_0 - 12 M^2 u^2_0}}{4M}, \\
	u_{2} &= u_0, \\
	u_{3} &= \frac{(a^2-2M u_0)+\sqrt{a^4 + 4M a^2 u_0 - 12 M^2 u^2_0}}{4M}.
\end{split}	
\label{eqn:u1u2u3}
\end{align}
It is straightforward to verify that when $\frac{1}{r_\Lambda}<u_0<\frac{a^2}{2M}$, all the three roots are real, with $u_1<0$ and $u_3\geq u_2>0$. Furthermore, the three roots can also be written in terms of $b_{\text{eff}}$ as
\begin{align}
	\begin{split}
		{u}_{1}   = \frac{ a^2 }{6 M}\left(1-2  \cos \frac{{\varsigma}}{3}\right), \qquad
		{u}_{2,3} = \frac{a^2}{6M} \left(1+\cos \frac{{\varsigma}}{3} \mp \sqrt{3}\sin \frac{{\varsigma}}{3}\right),
	\end{split}
	\label{eqn:u1u23}
\end{align}
with
$${\varsigma} \equiv \cos^{-1} \left(\frac{54M^2}{a^6 b_{\text{eff}}^2}-1\right).$$

In general, the positions of the observer $(r_O, \varphi_O)$ and the source $(r_S, \varphi_S)$ could be located on either the same or opposite sides of the turning point $r_0$. Here, we consider only the latter case, as shown in Fig.~\ref{fig:lens}, which is the conventional choice in the study of gravitational lensing. Equation (\ref{eqn:dudphi}) can be integrated after taking the square root:
\begin{align}
	\int^{\varphi}_{\varphi(u_0)}d\varphi' = \pm\int^{u}_{u_0} \frac{du'}{\sqrt{2M(u'-u_1)(u_2-u')(u_3-u')}},
	\label{eqn:intphi}
\end{align}
where the sign $+/-$ is chosen if the position $r$ is on the left/right side of $r_0$. By applying the integration formula \textbf{3.131.4} in Ref.~\cite{72}, we obtain the solution to Eq.~(\ref{eqn:dudphi}) as
\begin{align}
	\varphi(u)
	= \varphi(u_0) \pm \sqrt{\frac{2}{M(u_3-u_1)}} [\mathbf{K}(k) - \mathbf{F}(\beta(u),k)],
	\label{eqn:phiu}
\end{align}
where $\mathbf{F}(\beta,k)$ is the incomplete elliptic integral of the first kind,  $\mathbf{K}(k)=\mathbf{F}(\frac{\pi}{2},k)$ is the complete elliptic integral, and the parameter and argument are give by
\begin{align}
	k = \sqrt{\frac{u_2-u_1}{u_3-u_1}}, \qquad
	\beta(u) = \sin^{-1}\sqrt{\frac{u-u_1}{u_2-u_1}}.
	\label{eqn:kappabeta}
\end{align}
For completeness, we also provide the explicit expression of $u$ as a function of $\varphi$ by inverting Eq.~(\ref{eqn:phiu}):
\begin{align}
	u(\varphi)=u_1+(u_2-u_1)\text{sn}^2(q(\varphi),k),
	\label{eqn:uphi2}
\end{align}
where $\text{sn}(q,k)$ is the Jacobi elliptic function of the first kind, and
\begin{align}
	q(\varphi)=\mathbf{K}(k)-\sqrt{\frac{M(u_3-u_1)}{2}}|\varphi-\varphi(u_0)|. 
\end{align}

Denoting $u_O$ and $u_S$ as the inverse of the radial positions of the observer $r_O$ and the source $r_S$, and substituting Eq.~(\ref{eqn:phiu}) into Eq.~(\ref{eqn:alpha1beff}), we get
\begin{align}
	\alpha_{1D}(b_\text{eff})=\frac{2\left[2\mathbf{K}(k)- \mathbf{F}(\beta_O,k)-\mathbf{F}(\beta_S,k)\right]}{\sqrt{2M(u_3-u_1)}}
	\label{eqn:alpha1Db}
\end{align}
with $\beta_{O/S} \equiv \beta(u_{O/S})$. Finally, combining this first part of the deflection angle with the second part given in Eqs.~(\ref{eqn:alpha2}) and (\ref{eqn:Psis}), we obtain an analytical expression for the finite-distance deflection angle of light of the Schwarzschild--de Sitter--black-hole--global-monopole system as
\begin{align}
	\begin{split}
		\alpha_D(b,\Lambda) &= \frac{2\left[2\mathbf{K}(k) - \mathbf{F}(\beta_O,k) - \mathbf{F}(\beta_S,k)\right]}{\sqrt{2M(u_3-u_1)}} - \pi         
		 +\sin^{-1} \left[{u_O} b\sqrt{ \mathcal{A}_0(r_O)} \right] 
		 +\sin^{-1} \left[{u_S} b\sqrt{ \mathcal{A}_0(r_S)} \right].
	\end{split}
	\label{eqn:alphaDb}
\end{align}
This exact formula is the key result of this paper.

To verify this result, we first set $a=0$ in the absence of the global monopole in Eq.~(\ref{eqn:alpha1Db}), thereby recovering the results obtained in Ref.~\cite{27}. Furthermore, under the additional condition that the source and observer are very far, namely $r_{O/S}/M\gg 1$ but $r_{O/S}/r_\Lambda \ll 1$, the last two terms of Eq.~(\ref{eqn:alphaDb}) are negligible, leading to
\begin{align}
	\beta_{O/S}\to \bar{\beta} \equiv \sin^{-1} \sqrt{\frac{-u_1}{u_2-u_1}}.
	\label{eqn:betaSObeta2}
\end{align}
Thus, Eq.~(\ref{eqn:alphaDb}) becomes
\begin{align*}
	\alpha_D(b,\Lambda)\to \frac{4 [\mathbf{K}(k)-\mathbf{F}(\bar{\beta},k)]}{\sqrt{2M (u_3-u_1)}}-\pi.
\end{align*}
This is exactly the same expression for the deflection angle of light for the Schwarzschild black hole using Weinberg's method. Note that here $k$ and $\bar{\beta}$ implicitly depend on $b$, as the roots $u_i$ depend on $b$.

\section{Deflection Angles in the Weak-field Limit}
\label{sec:weak}

The reliance of the deflection angles in Eq.~(\ref{eqn:alphaDb}) on elliptical integrals hinders both its practical application and an intuitive grasp of its parameter dependencies. Therefore, approximate formulas for various limiting regimes are essential. We begin our analysis in the weak-field regime, deferring the strong-field limit to Secs.~\ref{sec:strong} and \ref{sec:observation}.

The weak-field deflection angle can be derived in two ways: by applying the weak-field approximation to the exact formula in Eq.~(\ref{eqn:alphaDb}), or by incorporating the approximation directly into the equations of motion before integration. Here, we adopt the former approach. Appendix \ref{sec:appendix} details the latter, providing an independent verification of our results within this asymptotically nonflat framework.

The weak-field limit corresponds to a light ray's closest approach, $r_0$, being significantly larger than the lens's characteristic gravitational size, $r_c$. In this regime, the light ray is only slightly deflected. Quantities then naturally fall into two groups with distinct magnitudes: $r_c,M\ll r_0,b,b_\text{eff}$.

By introducing the dimensionless parameter $\tilde{u} \equiv M u$, and noting that $\tilde{u}_0 \ll 1$, we can perturbatively expand the three roots in Eq.~(\ref{eqn:u1u2u3}) in powers of $\tilde{u}_0$. To fourth order of $\tilde{u}_0$, we find
\begin{align}
	\begin{split}
		&\tilde{u}_{1} \simeq - \tilde{u}_{0} + 2\frac{\tilde{u}^2_{0}}{a^2}  -4 \frac{\tilde{u}^3_{0}}{a^4} + 16 \frac{\tilde{u}^4_{0}}{a^6}  + \mathcal{O}\left( \tilde{u}_0^5\right), \\
		&\tilde{u}_{2} = \tilde{u}_{0}, \\
		&\tilde{u}_{3} \simeq \frac{a^2}{2} - 2\frac{\tilde{u}^2_{0}}{a^2} + 4\frac{\tilde{u}^3_{0}}{a^4} - 16 \frac{\tilde{u}^4_{0}}{a^6}
		  + \mathcal{O}\left( \tilde{u}_0^5\right) . 
	\end{split}	
	\label{eqn:utilde}
\end{align}
Substituting these into the expressions for $k$ and $\bar{\beta}$ in Eqs.~(\ref{eqn:kappabeta}) and (\ref{eqn:betaSObeta2}) yields:
\begin{align}
	\begin{split}
		k^2 \simeq 4\frac{\tilde{u}_{0}}{a^2} - 12\frac{\tilde{u}^2_{0} }{ a^4} + 64\frac{\tilde{u}^2_{0} }{a^6} - 320\frac{\tilde{u}^4_{0} }{a^8}  , \qquad
		\bar{\beta} \simeq \frac{\pi}{4} - \frac{\tilde{u}_{0}}{2 a^2} + \frac{\tilde{u}^2_{0}}{2 a^4} - \frac{31\tilde{u}^3_{0} }{12 a^6} + \frac{35\tilde{u}^4_{0} }{4 a^8} .
	\end{split}
\end{align}
Furthermore, $\beta_{O/S}$ can also be Taylor expanded in powers of $\tilde{u}_{O/S}$. Since  $\tilde{u}_{O/S} \ll \tilde{u}_0$ (because $r_{O/S} \gg r_0$), we only keep up to the third-order term:
\begin{align}
	\begin{split}
		\beta_{O/S} & \simeq \bar{\beta} + \left(\frac{1}{2 \tilde{u}_0} + \frac{1}{2 a^2 } - \frac{\tilde{u}_{0}}{4 a^4} + \frac{9\tilde{u}^2_{0}}{4 a^6} \right) \tilde{u}_{O/S} - \left(\frac{1}{4 a^2 \tilde{u}_0} + \frac{1}{4 a^4 }\right) \tilde{u}_{O/S}^2  + \left(\frac{1}{12 \tilde{u}_0^3} + \frac{1}{4 a^2 \tilde{u}_0^2} \right) \tilde{u}_{O/S}^3 \\
		&\quad
		+ \mathcal{O}\left(\tilde{u}_0^5, \tilde{u}_0^3\tilde{u}_{O/S}, \tilde{u}_0\tilde{u}_{O/S}^2, \frac{\tilde{u}_{O/S}^3}{\tilde{u}_0}, \frac{\tilde{u}_{O/S}^4}{\tilde{u}_0^3} , \frac{\tilde{u}_{O/S}^5}{\tilde{u}_0^5}\right) .
	\end{split}
\end{align}
Using the expansions of the elliptic integrals near $k=0$ (see \textbf{730.00}, \textbf{900.00}, and \textbf{902.00} in Ref.~\cite{elliptic}),
we obtain
\begin{align}
	\begin{split}
		\mathbf{F}(\beta_{O/S},k) &\simeq \mathbf{F}(\bar{\beta},k) + \left(\frac{1}{2\tilde{u}_0} + \frac{1}{a^2} - \frac{\tilde{u}_0}{a^4} + 6\frac{\tilde{u}_0^2}{a^6}\right) \tilde{u}_{O/S} + \left(\frac{1}{12\tilde{u}_0^3} + \frac{1}{3a^4\tilde{u}_0^2} \right) \tilde{u}_{O/S}^3, \\
		\mathbf{K}(k) &\simeq \frac{\pi}{2} + \frac{\pi}{2} \frac{\tilde{u}_{0}}{a^2} - \frac{3\pi}{8} \frac{\tilde{u}^2_{0}}{a^4} + \frac{35\pi}{8} \frac{\tilde{u}^3_{0}}{a^6} - \frac{1591\pi}{128} \frac{\tilde{u}^4_{0}}{a^8},
	\end{split}	
\end{align}
where
\begin{align*}
	\mathbf{F}(\bar{\beta},k) &\simeq \frac{\pi}{4} - \left(1-\frac{\pi}{4}\right) \frac{\tilde{u}_{0}}{a^2} - \frac{3\pi}{16} \frac{\tilde{u}^2_{0}}{a^4} - \left(\frac{14}{3} - \frac{35\pi}{16}\right) \frac{\tilde{u}_0^3}{a^6} + \left(\frac{16}{3} - \frac{1591\pi}{256}\right) \frac{\tilde{u}_0^4}{a^8}.
\end{align*}

Substituting these approximations into Eq.~(\ref{eqn:alpha1Db}), we derive the weak-field limit to the first part of the light deflection angle, up to fourth order in ${M}/{{r_0}}$:
\begin{align}
	\begin{split}
		\alpha_{1D}(b_\text{eff}) &\simeq \frac{\pi}{a} + \frac{4 M}{a^3r_0} - \left(4 - \frac{15\pi}{4}\right) \frac{M^2}{a^5 r^2_0}  + \left( \frac{122}{3} - \frac{15\pi}{2}\right) \frac{M^3}{a^7 r^3_0}  - \left( 130 - \frac{3465\pi}{64}\right) \frac{M^4}{a^9 r^4_0} \\
		&\quad - \left(\frac{r_0}{a} + \frac{M}{a^3} + \frac{3M^2}{2a^5 r_0} + \frac{5M^3}{2a^7 r_0^2}  \right) \left(\frac{1}{r_O}+\frac{1}{r_S}\right)
		- \left(\frac{r_0^3}{6a} + \frac{M r_0^2}{2a^3} 
		\right) \left(\frac{1}{r_O^3}+\frac{1}{r_S^3}\right) \\
		&\quad + \mathcal{O}\left( \frac{M^5}{r_0^5}, \frac{M^4}{r_0^3r_{O/S}}, \frac{M^3}{r_0r_{O/S}^2}, \frac{M^2r_0}{r_{O/S}^3},  \frac{Mr_0^3}{r_{O/S}^4},  \frac{r_0^5}{r_{O/S}^5}\right) .
	\end{split}
	\label{eqn:alpha1Dweak1}
\end{align}
For observational purposes, it is more convenient to express $r_0$ in terms of the (effective) impact parameter $b_\text{eff}$. Introducing a dimensionless parameter $\tilde{b} \equiv {b_\text{eff}}/{M} \gg 1$, and Taylor expand $\tilde{u}_0 = \tilde{u}_2$ from Eq.~(\ref{eqn:u1u23}) up to the fourth order in $1/\tilde{b}$, we find
\begin{align}
	\tilde{u}_0 &\simeq \frac{1}{a \tilde{b}} + \frac{1}{a^4\tilde{b}^2}  + \frac{5}{2a^7\tilde{b}^3} + \frac{8}{a^{10}\tilde{b}^4} + \mathcal{O}\left( \frac{1}{\tilde{b}^5}\right)  . 
	\label{eqn:utildeb}
\end{align}
Applying this approximation to Eq.~(\ref{eqn:alpha1Dweak1}), we express the first part of the deflection angle in terms of the impact parameter as
\begin{align}
	\begin{split}
	\alpha_{1D}(b_\text{eff}) &\simeq \frac{\pi}{a} + \frac{4 M}{a^4 b_\text{eff}} + \frac{15 \pi M^2}{4a^7 b_\text{eff}^2} + \frac{128 M^3}{3a^{10} b_\text{eff}^3} + \frac{3465 \pi M^4}{64a^{13} b_\text{eff}^4}  - b_\text{eff}\left(\frac{1}{r_O}+\frac{1}{r_S}\right)  - \frac{a^2b_\text{eff}^3}{6}\left(\frac{1}{r_O^3}+\frac{1}{r_S^3}\right)\\
	&\quad + \mathcal{O}\left( \frac{M^5}{b_\text{eff}^5}, \frac{M^4}{b_\text{eff}^3r_{O/S}}, \frac{M^3}{b_\text{eff}r_{O/S}^2}, \frac{M^2b_\text{eff}}{r_{O/S}^3},  \frac{Mb_\text{eff}^3}{r_{O/S}^4},  \frac{b_\text{eff}^5}{r_{O/S}^5}\right).	
	\end{split}
	\label{eqn:alpha1Dweak}
\end{align}
This result is consistent with that obtained by direct perturbative calculations using the Gauss-Bonnet theorem, as detailed in Appendix \ref{sec:appendix}.

Before proceeding to the second part of the deflection angle, it is convenient to reintroduce the cosmological constant. To the first order of $\Lambda$, Eq.~(\ref{eqn:beff}) gives
\begin{equation}
	b_{\text{eff}} \simeq b \left[1 - \frac{\Lambda b^2}{6} 
	+ \mathcal{O}\left( \Lambda^2 b^4 \right)\right]  .
	\label{eqn:beff2}	
\end{equation}
This holds if $b\ll {r_\Lambda}/{a}$, a condition easily satisfied in the weak-field limit. When $r_O$ and $r_S$ are much larger than the impact parameter $b$, but much smaller than $r_\Lambda$ (i.e., $r_\Lambda \gg r_{O/S}\gg b\sim r_0\gg  M $), we can expand the two inverse sine functions in the second part of the deflection angle as
\begin{align}
	\begin{split}
		\sin^{-1} \left[{u_O} b\sqrt{ \mathcal{A}(r_O)} \right] + \sin^{-1} \left[{u_S} b\sqrt{ \mathcal{A}(r_S)} \right] 
		&\simeq a b\left(\frac{1}{r_O}+\frac{1}{r_S}\right) -\frac{Mb}{a} \left(\frac{1}{r_O^2}+\frac{1}{r_S^2}\right) + \frac{a^3b^3}{6} \left(\frac{1}{r_O^3}+\frac{1}{r_S^3}\right) \\
		&\quad
		-\frac{\Lambda b}{6a} (r_O+r_S)
		- \frac{M\Lambda b}{3a^3} 
		- \left(ab^2 + \frac{3M^2}{a^5} \right)\frac{\Lambda b}{12} \left(\frac{1}{r_O} + \frac{1}{r_S}\right)\\
		&\quad 	+ \frac{M \Lambda b^3}{12a} \left(\frac{1}{r_O^2} + \frac{1}{r_S^2}\right) - \frac{a^3 \Lambda b^5}{16} \left(\frac{1}{r_O^3} + \frac{1}{r_S^3}\right) \\
		&\quad
		+ \mathcal{O}\left(  \frac{M^2b}{r_{O/S}^3}, \frac{Mb^3}{r_{O/S}^4}, \frac{b^5}{r_{O/S}^5},  \Lambda^2 b r_{O/S}^3 \right) .
	\end{split}	
\end{align}
It should be noted that this expansion differs from those in Refs.~\cite{24,24a,25,26,31}, where the assumption is $r_{O/S} \gg 2M $. In those references, $r_{O/S}$ could be of the same order as $b$ rather than $r_{O/S}\gg b$.

\begin{figure}[htp]
\centering
	\includegraphics[scale=1]{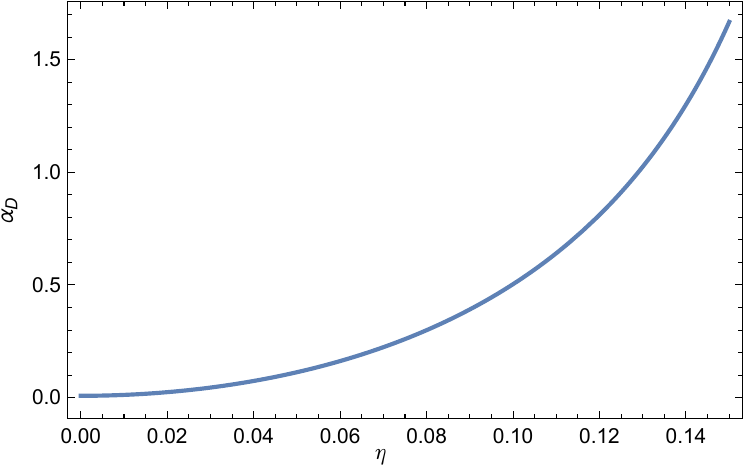}
   	\caption{The weak deflection angle $\alpha_D(b,\Lambda)$ in Eq.~(\ref{eqn:alphaDweak}) as a function of the gauge symmetry breaking parameter $\eta$. In this plot, the parameters are set to $M=1$, $b=500$, and $\Lambda=0$.}
   	\label{fig:weakfield1}
\end{figure}

By applying the approximation in Eq.~(\ref{eqn:beff2}) to the first part of the deflection angle in Eq.~(\ref{eqn:alpha1Dweak}) and combining it with the second part, we derive the weak-field limit for the deflection angle as
\begin{align}
	\begin{split}
		\alpha_D(b,\Lambda) 
		&\simeq
		(1-a)\frac{\pi}{a} + \frac{4 M}{a^4 b} 
		+ \frac{ 15 \pi M^2}{4a^7 b^2} + \frac{ 128 M^3}{3a^{10} b^3} + \frac{ 3465 \pi M^4}{64a^{13} b^4}
		-(1-a) b \left(\frac{1}{r_O}+\frac{1}{r_S}\right) - (1-a)a^2\frac{b^3}{6} \left(\frac{1}{r_O^3}+\frac{1}{r_S^3}\right)\\
		&\quad  - \frac{Mb}{a} \left(\frac{1}{r_O^2}+\frac{1}{r_S^2}\right) 
		- \frac{\Lambda b (r_O+r_S)}{6a}
		+ (2 - a)\frac{\Lambda b^3 }{12} \left(\frac{1}{r_O}+\frac{1}{r_S}\right)   + (4-3a)a^2\frac{ \Lambda b^5 }{48} \left(\frac{1}{r_O^3}+\frac{1}{r_S^3}\right)  \\
		&\quad + (2-a)\frac{M \Lambda b}{3 a^4 } + \frac{M \Lambda b^3 }{12a} \left(\frac{1}{r_O^2}+\frac{1}{r_S^2}\right) + \frac{5\pi M^2 \Lambda}{4 a^7 } - \frac{M^2 \Lambda b }{4a^5} \left(\frac{1}{r_O}+\frac{1}{r_S}\right) 	 \\
		&\quad 
		+ \mathcal{O}\left(   \frac{M^5}{b^5}, \frac{M^4}{b^3r_{O/S}}, \frac{M^3}{br_{O/S}^2},\frac{M^2b}{r_{O/S}^3},\frac{Mb^3}{r_{O/S}^4},\frac{b^5}{r_{O/S}^5}, \frac{M^3 \Lambda}{b},  \Lambda^2 b r_{O/S}^3 \right).
	\end{split}
	\label{eqn:alphaDweak}
\end{align}
It is evident that the terms involving $r_{O/S}$ represent the finite-distance correction, while the contribution of the global monopole to the deflection angle is encapsulated in $a$. For the physically significant case with a small value of $\eta$, the leading term is $(1-a)\pi/a\simeq 4\pi^2 \eta^2+ 24\pi^3 \eta^4$. As shown in Fig.~\ref{fig:weakfield1}, the deflection angle of light is enhanced due to the presence of the global monopole. This enhancement supports the hypothesis that the global monopole could serve as an alternative to dark matter. It is also noteworthy that the first two terms align with the results obtained in Ref.~\cite{60}, where the author calculated the deflection angle of light only up to the first order of ${M}/{b}$.

One way to verify our results is by comparing them with those in the literature. To do this, we can turn off the global monopole by setting $a=1$. In this case, the deflection angle in Eq.~(\ref{eqn:alphaDweak}) simplifies to
\begin{align*}
	\left.\alpha_D(b,\Lambda)\right|_{a=1}
	&\simeq \frac{4 M}{b} + \frac{15 \pi M^2}{4 b^2}  + \frac{128 M^3}{3 b^3} + \frac{3465 \pi M^4}{64 b^4} - Mb \left(\frac{1}{r_O^2}+\frac{1}{r_S^2}\right) - \frac{\Lambda b (r_O+r_S)}{6} + \frac{\Lambda b^3 }{12} \left(\frac{1}{r_O}+\frac{1}{r_S}\right)
	\\
	&\quad  
	  + \frac{ \Lambda b^5 }{48} \left(\frac{1}{r_O^3}+\frac{1}{r_S^3}\right)  + \frac{M \Lambda b}{3}  + \frac{M \Lambda b^3 }{12} \left(\frac{1}{r_O^2}+\frac{1}{r_S^2}\right) + \frac{5\pi M^2 \Lambda}{4} - \frac{M^2 \Lambda b }{4} \left(\frac{1}{r_O}+\frac{1}{r_S}\right) . 
\end{align*}
The first few terms are commonly found in all light deflection results. Several terms involving $r_{O/S}$ or $\Lambda$ have appeared in previous literature. For instance, the fifth, sixth, seventh, and ninth terms are reported in Ref.~\cite{19}. However, the coefficients of the fifth, seventh, and ninth terms in our expression are only half of those presented in Ref.~\cite{19}. Notably, the seventh and ninth terms also appear in Ref.~\cite{75}, again with coefficients that are half as large. We believe that this discrepancy arises from inconsistent asymptotic expansions in the previous works. 

As a final verification, we consider the scenario where the global monopole is absent ($a=1$) and the cosmological constant is zero ($\Lambda=0$). In this case, the weak deflection angle further reduces to
\begin{align*}
	\left.\alpha_D(b,\Lambda=0)\right|_{a=1} &\simeq \frac{4 M}{ b}+\frac{ 15 \pi M^2}{4b^2}  + \frac{128 M^3}{3 b^3} + \frac{3465 \pi M^4}{64 b^4} - Mb \left(\frac{1}{r_O^2}+\frac{1}{r_S^2}\right).
\end{align*}
This result aligns with the well-known deflection angle for the Schwarzschild black hole \cite{73,74}. The last term represents the finite-distance correction and is also obtained in Ref.~\cite{27}.

\section{Deflection Angles in the Strong-field Limit}
\label{sec:strong}

In the strong-field limit, the closest approach of a light ray is nearly at its gravitational radius ($r_0 \gtrsim r_c$) \cite{24a}. As $r_0 \to r_c$, the light deflection angle increases and eventually diverges. The deflection angle in this regime can be obtained by expanding the analytical expression from Eq.~(\ref{eqn:alphaDb}) near the photon sphere radius $r_c$. To facilitate this, we introduce a small parameter
\begin{align}
	\epsilon \equiv \frac{r_0-r_c}{M} \ll 1.
\end{align}
Using Eqs.~(\ref{eqn:beff}) and (\ref{eqn:rc}), we can expand the effective impact parameter $b_\text{eff}$ as
\begin{align}
	b_\text{eff} (r_0) \simeq b_\text{eff} (r_c) + \half \sqrt{3} aM \epsilon^2.
\end{align}
Recall that $b_\text{eff} (r_c) = 3\sqrt{3}M/a^3$, as given in Eq.~(\ref{eqn:beffrc}), represents the (effective) impact parameter when $r_0=r_c$. The three roots in Eq.~(\ref{eqn:u1u2u3}) can be Taylor-expanded around $b_\text{eff} (r_c)$, up to the second order, we have
\begin{align}
\begin{split}
	u_1 &\simeq  - \frac{1}{2} u_c  + \frac{2}{3} u^3_c M^2\epsilon^2,  \\
	u_2 &\simeq  u_c - u^2_c M \epsilon +  u^3_c M^2 \epsilon^2, \\
	u_3 &\simeq u_c +M u^2_c \epsilon - \frac{5}{3} u^3_c M^2\epsilon^2, 
\end{split}
\end{align}
where $u_c \equiv \frac{1}{r_c} = \frac{a^2}{3M}$. Substituting these expansions into Eq.~(\ref{eqn:kappabeta}) results in
\begin{align}
	k^2 &\simeq 1 - \frac{4}{3} u_c M \epsilon + \frac{8}{3} u^2_c M^2 \epsilon^2.
\end{align}
In the absence of a global monopole ($a=1$), the above equations reduce to those derived in Ref.~\cite{76}.

To derive the light deflection angle near the photon sphere, we apply the following formulas for $k \approxeq 1$ (refer to \textbf{112.01} and \textbf{111.04} in Ref.~\cite{elliptic}):
\begin{align*}
	\mathbf{K}(k)\simeq \frac{1}{2}\ln\left(\frac{16}{1-k^2}\right),\qquad
	\mathbf{F}(\beta,1) = \frac{1}{2}\ln\left(\frac{1+\sin\beta}{1-\sin\beta}\right).
\end{align*}
From these, we obtain
\begin{align}
\begin{split}
	\mathbf{K}(k) &\simeq -\frac{1}{2}\ln\left(\frac{u_c M\epsilon}{12}\right), \\
	\mathbf{F}(\beta_{O/S},k) &\simeq \frac{1}{2} \ln\left(\frac{1+\sqrt{\mathcal{D}_{O/S}}}{1-\sqrt{\mathcal{D}_{O/S}}}\right),
\end{split}
\end{align}
where
$$\mathcal{D}_{O/S} \equiv \frac{1}{3} + \frac{2 u_{O/S} M}{a^2}.$$
Substituting these results into Eq.~(\ref{eqn:alpha1Db}), and after some algebraic manipulations, we obtain the leading asymptotic behavior in the strong-field limit as
\begin{align}
	\alpha_{1D}(b_\text{eff}) \simeq - {A} \ln\left[ \frac{b_\text{eff} - b_\text{eff}(r_c)}{b_\text{eff}(r_c)}\right] + {B}_1,
	\label{eqn:alpha1Dstrong}
\end{align}
where
\begin{align}
	A   &\equiv \frac{1}{a} =\frac{1}{\sqrt{1-8\pi \eta^2}}, \nonumber\\
	B_1 &\equiv \frac{1}{a}\ln216 -\frac{1}{a}\ln\frac{(1+\sqrt{\mathcal{D}_O})(1+\sqrt{\mathcal{D}_S})}{(1-\sqrt{\mathcal{D}_O})(1-\sqrt{\mathcal{D}_S})}. \nonumber
\end{align}
If $r_{O/S}\gg M $, then $B_1 \to \frac{1}{a}\ln [216(7-4\sqrt{3})]$. The result in Eq.~(\ref{eqn:alpha1Dstrong}) is consistent with the findings in Ref.~\cite{62}. Explicitly expanding the above result to the leading order of the cosmological constant, we find
\begin{align}
	\alpha_{1D}(b,\Lambda) \simeq - {A} \ln\left( \frac{b-b_c}{b_c}\right) + {B}_1 +\frac{9M^2\Lambda}{a^7},
	\label{eqn:alpha1Dstrong2}
\end{align}
where we have used the fact that $ M^2 \Lambda\ll 1$.

A similar asymptotic analysis can be performed on the second part of the deflection angle. To the leading order in $\epsilon$ and $\Lambda$, we have
\begin{align}
	\alpha_{2D}(b,\Lambda)\simeq-\pi + C_{O}^{(0)} + C_{S}^{(0)}  + \left[C_{O}^{(1)} + C_{S}^{(1)}  \right]\Lambda,
	\label{eqn:alpha2Dstrong}
\end{align}
where
\begin{align}
C_{O/S}^{(0)} &\equiv \sin^{-1}\left[ \frac{b_c}{r_{O/S}} \sqrt{\mathcal{A}_0(r_{O/S})} \right] , \nonumber \\
C_{O/S}^{(1)} &\equiv - \frac{b_c r^2_{O/S}}{6 \sqrt{r^2_{O/S}\mathcal{A}_0(r_{O/S})- b^2_c \mathcal{A}^2_0(r_{O/S}) }}. \nonumber
\end{align}
\begin{figure}[htp]
	\centering
	\includegraphics[scale=1]{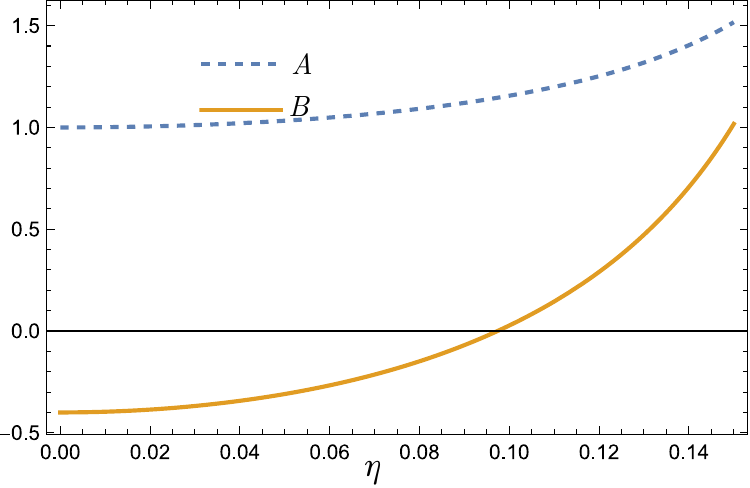}
	\caption{Dependence of the strong deflection coefficients $A$ (blue dashed line) and $B$ (orange solid line) in Eq.~(\ref{eqn:alphaDstrong}) on the gauge-symmetry breaking parameter $\eta$, where $0\leq\eta\leq 0.15$. Here, $M=1$ and $\Lambda=0$.}
	\label{fig:strongAB}
\end{figure}
Finally, combining Eqs.~(\ref{eqn:alpha1Dstrong2}) and (\ref{eqn:alpha2Dstrong}) yields the strong-field limit of the deflection angle of light as
\begin{align}
	\alpha_{D}(b,\Lambda) \simeq - {A} \ln\left( \frac{b-b_c}{b_c}\right)+{B},
	\label{eqn:alphaDstrong}
\end{align}
where
\begin{align}
	B \equiv {B}_1 -\pi + C_{O}^{(0)} + C_{S} ^{(0)} + \left[\frac{9M^2}{a^7} + C_{O}^{(1)} + C_{S}^{(1)} \right] \Lambda.
	\label{eqn:B}
\end{align}

We offer a few observations regarding the strong-field deflection angle. First, we plot the coefficients $A$ and $B$ from Eq.~(\ref{eqn:alphaDstrong}) as functions of the strength of the global monopole in Fig.~\ref{fig:strongAB}. The plot demonstrates that both coefficients are enhanced by the strength of the global monopole, which translates into an increased deflection angle. Similar to the weak-field limit, this enhancement supports the idea that the global monopole could be a possible alternative to dark matter. Second, note that $C_{O/S}^{(1)}<0$ and typically $r_{O/S}$ is much greater than $M$. In this case, the coefficient in front of $\Lambda$ in Eq.~(\ref{eqn:B}) is negative, indicating that the deflection angle of light is a decreasing function of the cosmological constant. 

In the absence of the global monopole, Eq.~(\ref{eqn:alphaDstrong}) recovers the result obtained in Ref.~\cite{27}. Finally, if $\Lambda=0$ and the source and observer are at the infinity, then Eq.~(\ref{eqn:B}) becomes
\begin{align}
	B\to\frac{1}{a}\ln [216(7-4\sqrt{3})] - \pi.
\end{align}

\section{Observables in Strong Lensing}
\label{sec:observation}

In this section, we apply the light deflection angle in the strong-field limit, as specified in Eq.~(\ref{eqn:alphaDstrong}), to calculate some observables in strong lensing. The exact lens equation on a flat background is derived in Ref.~\cite{78} and is uniformly extended to spherical and hyperbolic backgrounds in Ref.~\cite{79}. In the present work, the line element for the background is given by
\begin{align}
	ds^2=-A_\text{bg}(r)dt^2+\frac{dr^2}{A_\text{bg}(r)}+r^2 (d\theta^2+ \sin^2\theta d\varphi^2 ),
\end{align}
where the metric function is expressed as 
\begin{align*}
	A_\text{bg}(r) = a^2-\frac{\Lambda{r}^2}{3}=a^2(1-R^2),
\end{align*}
with $R\equiv{r}/{r_\Lambda}$. The optical metric for the background space, adhering to the condition $ds^2=0$, becomes
\begin{align}
	d\ell^2= \frac{3}{\Lambda}\left[\frac{dR^2}{1-R^2}+\frac{R^2}{1-R^2} (d\theta^2+ \sin^2\theta d\varphi^2 )\right].
\end{align}
From this, we can derive the Gauss curvature $K^\text{opt}=-a^4/r^{2}_\Lambda$, indicating a three-dimensional space of constant negative curvature. In this background, the lens equation in the strong-field limit is
\begin{align}
\begin{split}
	&\alpha_D - \sin^{-1}\left( \sqrt{\frac{1+K^\text{opt} {D}^2_{OS} \tan^2 \varrho}{{D}^2_{LS} +{D}^2_{OS} \tan^2 \varrho }} {D}_{OL}\sin \vartheta\right) -\vartheta +\tan^{-1}\left( \frac{{D}_{OS}}{{D}_{LS}}\tan \varrho\right) =2n\pi,
\end{split}	
\label{eqn:lenseqn}
\end{align}
where $n$ is the winding number for the light ray, $\varrho$ is the angular separation between the source and the lens, $\vartheta$ is the angular separation between the lens and the image, and $D_{AB}$ represents the physical distances between points $A$ and $B$ in this hyperbolic background. In this context, $L$ denotes the lens, $O$ is the observer, $S$ is the source, and
$$D_{{OS}} = D_{{OL}} + D_{LS}.$$

In practice, all light reaching us as observers in the solar system has a small deflection angle modulus to $2\pi$. For simplicity, consider relativistic images with rays winding clockwise around the black hole. We can express $\alpha_D=2n\pi+\Delta \alpha_n$, with $n$ an integer and $0<\Delta \alpha_n \ll 1$. We also assume small angles, with $|\varrho|\ll 1$ and $|\vartheta|\ll 1$, so we have $\tan \varrho\simeq \varrho$ and $\sin \vartheta \simeq \vartheta$. Furthermore, from the relation between the impact parameter of a light ray $b$ and the image position $D_{OL}$, given by $b=D_{OL}\sin \vartheta$, we find
\begin{align}
	b\simeq D_{OL}\vartheta.
	\label{eqn:bDtheta}
\end{align}
Note that $\varrho$ is very small. Thus, it can be expressed using a bookkeeping expansion parameter $\varepsilon$ as \cite{80}
\begin{align}
	\varrho= \varepsilon\varrho^{(1)}.
\end{align}
On the other hand, small $\vartheta$ and $\Delta \alpha_n$ can be considered as Taylor expansions started from the first order in $\varepsilon$:
\begin{align}
	\vartheta=\sum^\infty_{m=1} \vartheta^{(m)} \varepsilon^m, \qquad
	\Delta \alpha_n=\sum^\infty_{m=1} \Delta \alpha_n^{(m)} \varepsilon^m,
\end{align}
where $\vartheta^{(m)}$ and $\Delta \alpha_n^{(m)}$ are the corresponding coefficients. Substituting these expansions into Eq.~(\ref{eqn:lenseqn}), at the first order of $\varepsilon$, we obtain the linearized lens equation:
\begin{align}
	\varrho^{(1)}=\vartheta^{(1)}-\frac{D_{LS}}{D_{OS}} \Delta \alpha_n^{(1)}.
	\label{eqn:lenseqnlinear}
\end{align}
In the following, we will consider only this first-order lens equation. By multiplying both sides by $\varepsilon$, we obtain the same form as presented in Refs.~\cite{81,82}.

Solving Eq.~(\ref{eqn:alphaDstrong}) for the impact parameter $b$ and applying Eq.~(\ref{eqn:bDtheta}), we find the expression for the angle $\vartheta$ as
\begin{align}
	\vartheta \simeq \frac{b_c}{D_{OL}}\left[1+\exp{\left(\frac{B-\alpha_D}{A} \right)}\right].
\end{align}
The position of the $n$th relativistic image can be approximated by a first-order Taylor expansion around $\alpha_D=2 n \pi$ as
\begin{align}
	\vartheta_n \simeq \vartheta^0_n-\xi_n \Delta \alpha_n,
	\label{eqn:nthimage}
\end{align}
where
\begin{align}
\begin{split}
	\vartheta^0_n &\equiv \vartheta|_{\alpha_D=2n\pi} \simeq \frac{b_c(1+e_n)}{D_{OL}}, \\
	\xi_n &\equiv - \left.\frac{d\vartheta}{d\alpha_D}\right|_{\alpha_D=2n\pi} \simeq  \frac{b_c e_n}{AD_{OL}}, 
\end{split}
\end{align}
with
\begin{align*}
	e_n \equiv \exp \left(\frac{B-2n\pi}{A}\right).
\end{align*}
One can then approximately obtain the position of the $n$th relativistic image as \cite{78}
\begin{align}
	\vartheta_n\simeq \vartheta^0_n+\frac{b_c e_n (\varrho-\vartheta^0_n)D_{OS}}{{A} D_{OL}D_{LS}}.
\end{align}

\begin{figure}[htp]
	\centering
	\includegraphics[scale=1]{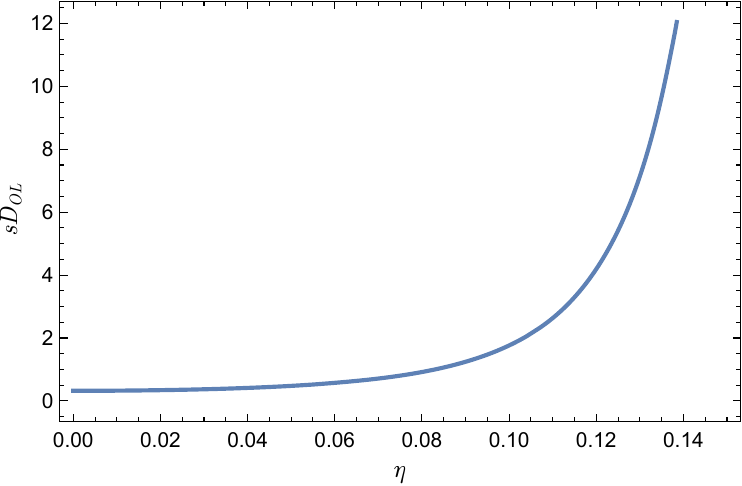}
	\caption{Angular separation $s$ between the outermost and asymptotic images as a function of  $\eta$, expressed in units of $D_{OL}$. Here, $0\leq\eta\leq 0.15$ and $M=1$. }
	\label{fig:s-eta}
\end{figure}

By taking the limit as $n\rightarrow \infty$ in Eq.~(\ref{eqn:nthimage}), we obtain the asymptotic angular position of the images:
\begin{align}
	\vartheta_\infty=\frac{b_c}{D_{OL}}.
\end{align}
Thus, the angular separation between the outermost and asymptotic images is given by
\begin{align}
	s \equiv \vartheta_1-\vartheta_\infty=\vartheta_\infty \exp \left(\frac{{B}-2\pi}{{A}}\right).
	\label{eqn:s}
\end{align}
In Fig.~\ref{fig:s-eta}, we plot this separation as a function of the global monopole strength $\eta$, and observe that the global monopole enhances the light deflection effect.

\begin{figure}[htp]
	\centering
	\includegraphics[scale=1]{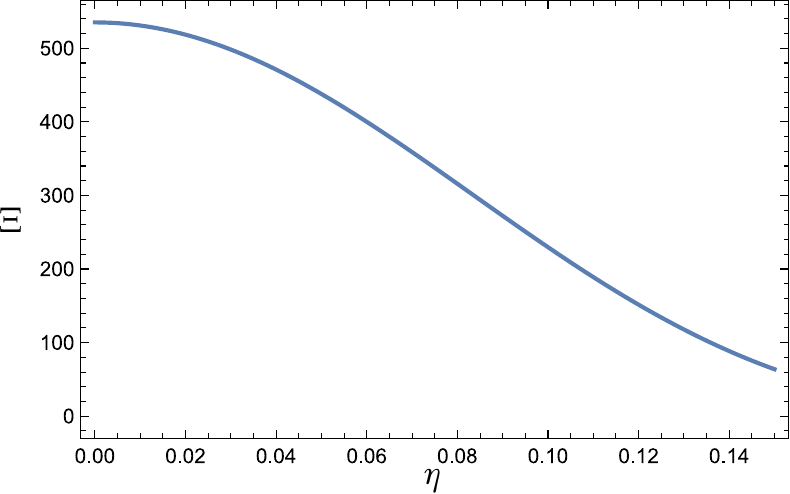}
	\caption{The relativistic magnification $\Xi$ of the outermost image as a function of $\eta$ is shown for the range $0\leq\eta\leq 0.15$. In this figure, we set $M=1$.}
	\label{fig:r-eta}
\end{figure}

The magnification of the $n$th image is given by \cite{78,83,84}
\begin{align}
	\mu_n \equiv \left| \frac{\varrho}{\vartheta} \frac{d\varrho}{d\vartheta}\right|^{-1}_{\vartheta^0_n}
\end{align}
In our context, it has the leading behavior as 
\begin{align}
	\mu_n \simeq \frac{b^2_c e_n (1+e_n)D_{OS}}{|\varrho|{A} D^2_{OL}D_{LS}}.
	\label{eqn:mun}
\end{align}
The relativistic magnification of the outermost image is defined as
\begin{align}
	\Xi \equiv \frac{\mu_1}{\sum^\infty_{n=2}\mu_n}.
\end{align}
Using Eq.~(\ref{eqn:mun}), we simplify the magnification to the following form:
\begin{align}
	\Xi \simeq \frac{\left(1+e^{\frac{{B}-2\pi}{{A}}}\right) \left(e^{\frac{4\pi}{{A}}}-1\right)}{e^{\frac{2\pi}{{A}}}+1+e^{\frac{{B}-2\pi}{{A}}} }.
	\label{eqn:r}
\end{align}
We plot $\Xi$ as a function of $\eta$ in Fig.~\ref{fig:r-eta} and observe that the presence of a global monopole suppresses the magnification. In the absence of both the global monopole and cosmological constant ($\Lambda=0$, $\eta=0$), and taking the large $r_{O/S}$ limit, we note that $e^{\frac{2\pi}{{A}}}\gg 1$, and that $e^{\frac{{B}}{{A}}}$, $A$, and $B$ are all of order unity. Under these conditions, Eq.~(\ref{eqn:r}) can be approximated as $\Xi\approx e^{\frac{2\pi}{A}}$, consistent with Ref.~\cite{78}. Furthermore, from Eqs.~(\ref{eqn:s}) and (\ref{eqn:r}), it is, in principle, possible to express the strong deflection coefficients $A$ and $B$ in terms of $s$ and $\chi$, thereby enabling the reconstruction of the full strong-field limit expansion of the deflection angle for the observed gravitational lens.

\section{Summary}
\label{sec:summary}

In this paper, we derived an exact analytical expression for the finite-distance deflection angle of light in the Schwarzschild--de Sitter black hole spacetime with a global monopole, using elliptic integrals. Based on this expression, we analyzed gravitational lensing in both the weak- and the strong-field limits. Our findings recover various results from the literature when the global monopole is absent. In the weak-field limit, we determined the deflection angle using Taylor expansions of the elliptic integrals, confirmed independently by direct perturbation calculations and the Gauss-Bonnet theorem. We explicitly demonstrated the contributions of finite-distance correction, the global monopole strength, and the cosmological constant, finding that the deflection angle is enhanced by a global monopole and increases with the parameter $\eta$. 

In the strong-field limit, the deflection angle is expressed through coefficients $A$ and $B$, with $A$ depending solely on the monopole and $B$ found to decrease with the cosmological constant $\Lambda$. We calculated observables such as the angular separation  $s$ between the outermost and asymptotic relativistic images, and the relativistic magnification $\Xi$ of the outermost image, using the linearized lens equation. Both strong lensing coefficients $A$ and $B$, as well as the angular separation $s$, increase monotonically with $\eta$. The enhancement of light deflection in both weak- and strong-field limits further strengthens the case for the global monopole as an alternative of dark matter.

\begin{acknowledgments}
M.Y.L.~is supported by the National Natural Science Foundation of China with Grant No.~12305064 and Jiangxi Provincial Natural Science Foundation with Grant No.~20224BAB211020. Q.-h.W.~thanks Mr.~Wu Quanrun for pointing out errors in Figs.~\ref{fig:weakfield1} and \ref{fig:strongAB}.
\end{acknowledgments}

\appendix

\section{Direct Perturbative Calculation of $\alpha_{1D}$ Using the Gauss-Bonnet Theorem without Cosmological Constant}
\label{sec:appendix}

\begin{figure}[htp]
	\centering
	\includegraphics[scale=0.65]{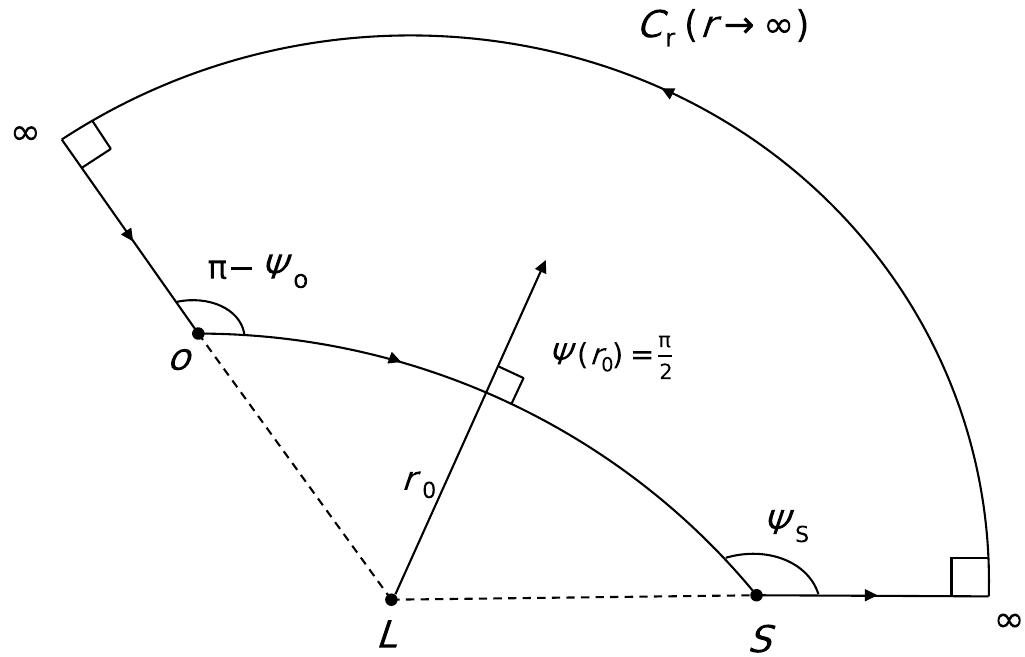}
	\caption{Schematic illustration of the integration domain: The quadrilateral ${}^\infty_O \Box_S^\infty$ is embedded in an asymptotic nonflat spacetime, similar to Fig.~2 in Ref.~\cite{31}. In this figure, $C_r$ represents a light ray path extending from the observer ($O$) to the source ($S$), and $C_\infty$ is a circular arc segment with a radius $R \to \infty$. The angular coordinate is set to $\frac{\pi}{2}$ at the point of closest approach, $r_0$.}
	\label{fig:quadrilateral}
\end{figure}

To further verify our results in this asymptotically nonflat background, we conduct a direct perturbative calculation of $\alpha_{1D}$ in the weak-field limit by applying the Gauss-Bonnet theorem:
\begin{align}
	\iint_T K dS + \int_{\partial T} \kappa_g d\ell + \sum^N_i \theta_i = 2\pi,
\end{align}
where $K$ denotes the Gaussian curvature of the orientable surface $T$, which has a boundary $\partial T$ consisting of differentiable curves with geodesic curvature $\kappa_g$. The terms $\theta_i$ represent the jump angles between the curves, $dS$ is the area element of the surface, and $\ell$ is the line element along the boundary. In our setup, the integration domain is depicted in Fig.~\ref{fig:quadrilateral}. Within this configuration, the Gauss-Bonnet theorem yields \cite{85,86}
\begin{align}
\begin{split}
	&\iint_{ ^\infty_O \Box_S^\infty} K dS + \int^O_{r_\infty}{\kappa}_g d\ell + \int^{r_\infty}_S{\kappa}_g d\ell + \int_O^S {\kappa}_g d\ell  + \int_{C_\infty}\kappa_g d\ell + \Psi_O + (\pi-\Psi_S) + \pi = 2\pi,
\end{split}
\label{eqn:Gauss-Bonnet}
\end{align}
where the quadrilateral ${}^\infty_O \Box_S^\infty$ comprises the photon orbit extending from the observer to the source in a generalized optical metric. It also includes two radial lines originating from $O$ and $S$, and $C_\infty$, a circular arc segment with a radius $R\gg r_{O/S}$ \cite{85}.

In the spacetime described in Eq.~(\ref{eqn:ds2}), the optical metric on the equatorial plane reads
\begin{align}
	dt^2=\gamma_{IJ}dx^I dx^J=\frac{1}{\mathcal{A}^2_0(r)}dr^2+\frac{r^2}{\mathcal{A}_0(r)}d\varphi^2,
	\label{eqn:ds2optical}
\end{align}
with the nonvanishing components of the metric given by
\begin{align}
	\gamma_{rr} &= \frac{1}{\mathcal{A}_0^2(r)} = \frac{1}{\left(a^2 -\frac{2M}{r}\right)^2}, \nonumber \\
	\gamma_{\varphi\varphi} &= \frac{r^2}{\mathcal{A}_0(r)}=\frac{r^2}{a^2 -\frac{2M}{r}}. \nonumber
\end{align}

Using the Liouville formula, we calculate the geodesic curvature $\kappa_g$ for a circular arc segment $C_R$ with $r(\varphi) = R = \text{const}$ as follows \cite{87}:
\begin{align}
	\kappa_g \big|_{r=R} = \left. \frac{1}{2\sqrt{\gamma_{rr}}}\frac{\partial \ln \gamma_{\varphi\varphi} }{\partial r} \right|_{r=R} =\frac{a^2}{R}-\frac{3M}{R^2}.
	\label{eqn:kappa_g}
\end{align}
Meanwhile, for a constant $R$, the optical metric in Eq.~(\ref{eqn:ds2optical}) yields
\begin{align}
	d\ell=\frac{R}{\sqrt{\mathcal{A}_0(R)}}d\varphi.
\end{align}
Since $R$ is very large, we find that $\kappa_g(C_R) d\ell \rightarrow a d\varphi $. Therefore, we have
\begin{align}
	\int_{C_\infty}\kappa_g d\ell=a\, \varphi_{OS}.
\end{align}
Similarly, for the radial lines where $\varphi=\varphi(r)= \text{const}$, we get
\begin{align}
	\kappa_g \big|_{\varphi=\text{const}} = \left. - \frac{1}{2\sqrt{\gamma_{\varphi\varphi}}}\frac{\partial \ln \gamma_{rr} }{\partial \varphi} \right|_{\varphi=\text{const}} = 0.
\end{align}
The geodesic curvature along the light ray from the observer to the source in the generalized optical metric given by Eq.~(\ref{eqn:ds2optical}) is also found to be zero:
\begin{align}
	\int_O^S {\kappa}_g d\ell =0.
\end{align}
Therefore, Eq.~(\ref{eqn:Gauss-Bonnet}) simplifies to
\begin{align}
	\iint_{ ^\infty_O \Box_S^\infty} K dS + a \varphi_{OS} + \Psi_O - \Psi_S = 0.
\end{align}
The first part of the deflection angle of light can then be derived from the above equation as
\begin{align}
\begin{split}
	\alpha_{1D}
	&= \varphi_{OS} \\
	&= \frac{\pi}{a} - \frac{1}{a}\iint_{ ^\infty_O \Box_S^\infty} K dS -\frac{1}{a}\sin^{-1} \left[u_O b\sqrt{\mathcal{A}_0(r_O)}\right]  -\frac{1}{a}\sin^{-1} \left[u_Sb\sqrt{\mathcal{A}_0(r_S)}\right].
\end{split}
\label{eqn:alpha1DGauss}
\end{align}

The Gaussian curvature $K$ can be calculated from the Riemann curvature tensor $R_{r\varphi r\varphi}$ as
\begin{align}
\begin{split}
	K &= \frac{R_{r\varphi r\varphi}}{\text{det}(\gamma_{IJ})} \\
	&=\frac{1}{\sqrt{\text{det}(\gamma_{IJ})}} \left[{\partial_\varphi}\left(\frac{\partial_\varphi \sqrt{\gamma_{rr}}} {\sqrt{\gamma_{\varphi\varphi}}}\right)- {\partial _r}\left( \frac{\partial_r \sqrt{\gamma_{\varphi\varphi}}} {\sqrt{\gamma_{rr}}}   \right)\right] \\
	&=-\frac{2a^2 M}{r^3}+\frac{3 M^2}{r^4}.
	\label{eqn:K}
\end{split}
\end{align}
The area element on the equatorial plane is given by
\begin{align}
\begin{split}
	dS &= \sqrt{\text{det}(\gamma_{IJ})} \, drd\varphi \\
	&=r \left(a^2 -\frac{2M}{r}\right)^{-{3}/{2}}drd\varphi.
\end{split}
\end{align}
In the weak-field limit $M\ll r$, this can be approximated as
\begin{align}
	dS \simeq \frac{r}{a^3}\left(1+\frac{3M}{a^2 r} + \frac{15M^2}{2a^4 r^2}  + \frac{35M^3}{2a^6 r^3}  + \frac{315M^4}{8a^8 r^4}\right) drd\varphi.
\end{align}
Substituting these expressions into the surface integration of the Gaussian curvature, we obtain
\begin{align}
\begin{split}
	-\iint K dS
	&\simeq \int^{{\varphi_O}}_{\varphi_S}d\varphi \int^\infty_{r( \varphi)} dr \left(\frac{2M}{ar^2} + \frac{3 M^2}{a^3 r^3} + \frac{6 M^3}{a^5 r^4} + \frac{25 M^4}{2a^7 r^5} \right) \\
	&\simeq \int^{{\varphi_O}}_{\varphi_S} d\varphi \int_0^{u(\varphi)} du \left(\frac{2M}{a} + \frac{3 M^2}{a^3} u + \frac{6 M^3}{a^5} u^2 + \frac{25 M^4}{2a^7} u^3 \right)  \\
	&= \int^{{\varphi_O}}_{\varphi_S} d\varphi\left[\frac{2M}{a} u(\varphi) + \frac{3M^2}{2a^3} u^2(\varphi) + \frac{2M^3}{a^5} u^3(\varphi) + \frac{25M^4}{8a^7} u^4(\varphi)\right].
\end{split}
\label{eqn:surfaceint}
\end{align}

To proceed, let us shift the variable $\varphi$ such that $r(\frac{\pi}{2}) = r_0$. Since $r_0$ represents the minimal distance, we immediately have $r'({\frac{\pi}{2}})=0$. Next, we expand $u(\varphi)$ in terms of $M/b$ (for simplicity, we disregard the difference between $b$ and $b_\text{eff}$ in this appendix) as follows:
\begin{align}
	\begin{split}
		u(\varphi) &\simeq u^{(0)}(\varphi)+\frac{M}{b} u^{(1)}(\varphi) + \frac{M^2}{b^2} u^{(2)}(\varphi)   + \frac{M^3}{b^3} u^{(3)}(\varphi) + \frac{M^4}{b^4} u^{(4)}(\varphi),
	\end{split}
	\label{eqn:uphi}
\end{align}
where
\begin{align*}
	u^{(0)}(\varphi)=\frac{\sin (a \bar{\varphi})}{ab}
\end{align*}
and
$$\bar{\varphi} \equiv \varphi+\frac{(1-a)\pi}{2a},$$
which represents the inverse of the orbital motion in the absence of the black hole ($M=0$). Furthermore,
\begin{align}
\begin{split}
	u^{(1)}(\varphi) &= \frac{1+ \cos^2 (a \bar{\varphi})}{a^4b}, \\
	u^{(2)}(\varphi) &= \frac{\cos (a \bar{\varphi}) \left[ 15  (\pi-2a\bar{\varphi})+{20 \tan (a\bar{\varphi}) - 3\sin(2a\bar{\varphi})}\right]}{8a^7 b}.\\
	&\vdots
\end{split}
\label{eqn:uphi012}
\end{align}
It is apparent that the form of $u(\varphi)$ reduces to the textbook case when there is no global monopole ($a=1$).

The leading term $u^{(0)}(\varphi)$ is derived by considering the pure effect of the global monopole, which means to set $M=0$. In this case, the spacetime is asymptotically nonflat, yet we can still assume that $r_{O/S}\rightarrow \infty$. Therefore, we find
\begin{align}
	\varphi_O= \frac{(a+1)\pi}{2a}, \qquad \varphi_S=\frac{(a-1)\pi}{2a}.
\end{align}
It is noteworthy that the difference $\varphi_O-\varphi_S=\frac{\pi}{a}$, which deviates from $\pi$ due to the presence of the global monopole.

In the presence of the Schwarzschild black hole, we assume the following adjustments:
\begin{align}
\begin{split}
	\varphi_O = \frac{(a+1)\pi}{2a}-\Delta \varphi_O, \qquad
	\varphi_S = \frac{(a-1)\pi}{2a}+\Delta \varphi_S, 
\end{split}
\end{align}
where $\Delta \varphi_{O/S}\ll 1$. Consequently, Eq.~(\ref{eqn:uphi}) can be rewritten as
\begin{align}
	\frac{1}{r_S} &= \frac{\sin (a \Delta\varphi_S)}{ab}+ \frac{M[1+ \cos^2 (a \Delta\varphi_S)]}{a^4b^2} + \frac{M^2 \cos (a \Delta\varphi_S)}{8a^7 b^3} 
 	\left[ 15  (\pi-2a\Delta\varphi_S)+{20 \tan (a\Delta\varphi_S) - 3\sin(2a\Delta\varphi_S)}\right] +\cdots.
\end{align}
Approximating to the third order in $\Delta \varphi_S$, we find
\begin{align}
	\begin{split}
		\frac{1}{r_S} &\simeq \frac{2M}{a^4b^2} + \frac{15\pi M^2}{8a^7b^3} + \frac{20 M^3}{a^{10}b^4} + \frac{2985\pi M^4}{128a^{13}b^5}  + \left(\frac{ 1}{b} - \frac{2M^2}{a^6 b^3}  - \frac{15\pi M^3}{4a^9 b^4} \right) \Delta\varphi_S 
		- \left(\frac{M}{a^2 b^2} + \frac{15\pi M^2 }{16 a^5 b^3}\right) (\Delta\varphi_S)^2 - \frac{a^2}{6 b} (\Delta\varphi_S)^3.
	\end{split}
\end{align}
The approximate solution for $\Delta \varphi_S$ depends on the relative magnitudes of the two independent small parameters, $M/b$ and $b/r_S$. Expanding up to fourth order in $M/b$ and  third order in $b/r_S$, we obtain
\begin{align}
	\begin{split}
		\Delta \varphi_S &\simeq \frac{b}{r_S} + \frac{a^2 b^3}{6 r_S^3} - \frac{2M}{a^4 b} - \frac{15 \pi M^2}{8a^7 b^2} - \frac{64 M^3}{3 a^{10} b^3} - \frac{3465 \pi M^4}{128a^{13} b^4}.
	\end{split}
	\label{eqn:DphiS}
\end{align}
Similarly, for the observer side, we get
\begin{align}
	\begin{split}
		\Delta \varphi_O &\simeq \frac{b}{r_O} + \frac{a^2 b^3}{6 r_O^3} - \frac{2M}{a^4 b} - \frac{15 \pi M^2}{8a^7 b^2} - \frac{64 M^3}{3 a^{10} b^3} - \frac{3465 \pi M^4}{128a^{13} b^4}.
	\end{split}
	\label{eqn:DphiO}
\end{align}

Finally, we are prepared to perform the surface integration in Eq.~(\ref{eqn:surfaceint}). We find
\begin{align}
\begin{split}
	-\iint K dS
	&\simeq\frac{2 M [\cos (a\Delta\varphi_O) +\cos (a\Delta\varphi_S )]}{a^3 b}  + \frac{15  M^2(\varphi_O-\varphi_S)}{ 4 a^5 b^2}  +\frac{M^2 [\sin (2a\varphi_O) - \sin (2a\varphi_S)]}{8a^6 b^2} +\cdots \\
	&\simeq \frac{4 M}{a^3 b} + \frac{15\pi  M^2}{4 a^6 b^2}  + \frac{128  M^3}{3 a^9 b^3}  +\frac{3465\pi  M^4}{64 a^{12} b^4} 
	 - \frac{Mb}{a}\left(\frac{1}{r^2_O}+\frac{1}{r^2_S}\right).
\end{split}
\end{align}
Substituting this into Eq.~(\ref{eqn:alpha1DGauss}) gives
\begin{align}
\begin{split}
	\alpha_{1D} &\simeq \frac{\pi}{a} + \frac{4 M}{a^4 b} + \frac{15\pi  M^2}{ 4 a^7 b^2} + \frac{128  M^3}{3 a^{10} b^3}  +\frac{3465\pi  M^4}{64 a^{13} b^4} 
	- \frac{Mb}{a^2}\left(\frac{1}{r^2_O}+\frac{1}{r^2_S}\right)\\
	&\quad -\frac{1}{a}\sin^{-1} \left[u_Sb\sqrt{\mathcal{A}_0(r_S)} \right] -\frac{1}{a}\sin^{-1} \left[u_O b\sqrt{\mathcal{A}_0(r_O)} \right] \\
	&\simeq \frac{\pi}{a}+\frac{4 M}{a^4 b}+\frac{15\pi  M^2}{ 4 a^7 b^2} + \frac{128  M^3}{3 a^{10} b^3}  +\frac{3465\pi  M^4}{64 a^{13} b^4}  - b\left(\frac{1}{r_O}+\frac{1}{r_S}\right) - \frac{a^2b^3}{6}\left(\frac{1}{r^3_O}+\frac{1}{r^3_S}\right).
\end{split}	
\end{align}
This result matches Eq.~(\ref{eqn:alpha1Dweak}) with $b_\text{eff}$ replaced by $b$, thus providing an independent verification in this complex setup. Interestingly, the surface integration can be bypassed by directly computing the difference of angles:
\begin{align}
\begin{split}
	\alpha_{1D} &= \varphi_O-\varphi_S \\
	&\simeq \frac{\pi}{a}-\Delta\varphi_O-\Delta\varphi_S \\
	&\simeq \frac{\pi}{a}+\frac{4 M}{a^4 b}+\frac{15\pi  M^2}{ 4 a^7 b^2} + \frac{128  M^3}{3 a^{10} b^3}  +\frac{3465\pi  M^4}{64 a^{13} b^4}  - b\left(\frac{1}{r_O}+\frac{1}{r_S}\right) - \frac{a^2b^3}{6}\left(\frac{1}{r^3_O}+\frac{1}{r^3_S}\right).
\end{split}
\end{align}
This shortcut further confirms the consistency of our perturbative calculations.

\section{Calculation of $\alpha_{D}$ using the Gauss-Bonnet Theorem in Schwarzschild--de Sitter Spacetime}
\label{sec:appendixB}

In the presence of a cosmological constant $\Lambda$, the radial coordinate $r$ cannot be taken to infinity, and the Gauss-Bonnet method presented in Appendix~\ref{sec:appendix} is therefore not applicable. Fortunately, a new Gauss-Bonnet approach for asymptotically nonflat spacetimes and finite-distance gravitational lensing has been developed in Ref.~\cite{25}. In this section, we apply this method to calculate the weak-field limit of gravitational deflection by a global monopole in Schwarzschild-–de Sitter spacetime. As in Sec.~\ref{sec:weak}, we employ the weak-field approximation, assuming $r_\Lambda \gg r_{O/S} \gg b \sim r_0 \gg M$. In this scenario, the optical metric on the equatorial plane is given by replacing the function $\mathcal{A}_0(r)$ in Eq.(\ref{eqn:ds2optical}) with $\mathcal{A}(r)$ as defined in Eq.(\ref{eqn:metricA}):
\begin{align}
	\mathcal{A}(r)=a^2 -\frac{2M}{r}-\frac{\Lambda r^2}{3}.
	\label{eqn:opticalmetricA}
\end{align}

\begin{figure}[htp]
	\centering
	\includegraphics[scale=1]{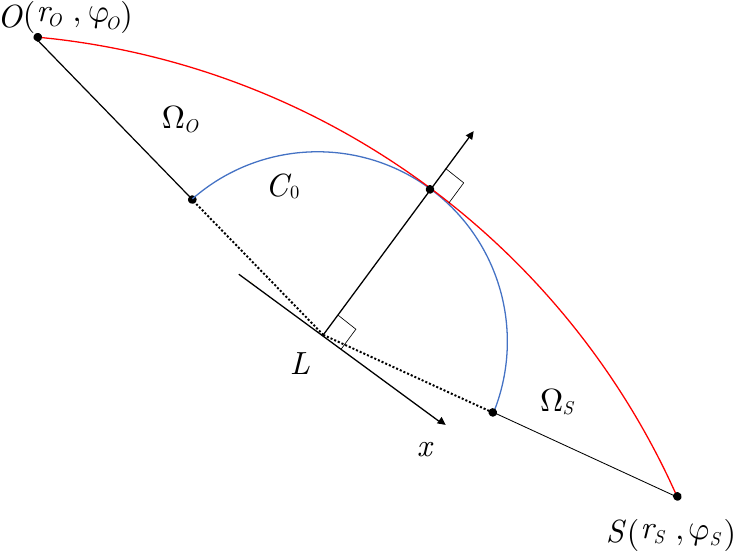}
	\caption{Schematic illustration of the integration domain considered in this appendix. This figure is reproduced from Fig.~5 in Ref.~\cite{25}.}
	\label{fig:GB2}
\end{figure}

Following Sec.~II.~C of Ref.~\cite{25}, we consider a quadrilateral bounded by the actual light path, a circular arc segment tangent to the light path at constant radial coordinate [$C_0: r(\varphi) = r_0$], and two radial paths connecting these two curves, as illustrated in Fig.~\ref{fig:GB2}. The geodesic curvature of the circular arc can be calculated using the Liouville formula \cite{76}. Remarkably, the dependence on $\Lambda$ cancels out, yielding the same result as in Eq.~(\ref{eqn:kappa_g}), with $R$ replaced by $r_0$: 
\begin{align}
	\kappa_g(r_0)=\frac{1}{2\sqrt{\gamma_{rr}}}\frac{\partial \ln \gamma_{\varphi\varphi} }{\partial r}=\frac{a^2}{r_0}-\frac{3M}{r_0^2},
\end{align}
where $r_0$ is defined by Eq.~(\ref{eqn:beff}). Similarly, the infinitesimal distance of $C_r$ reads 
\begin{align}
	d\ell=\frac{r_0}{\sqrt{\mathcal{A}(r_0)}}d\varphi.
\end{align}

The deflection angle can be expressed as \cite{25} 
\begin{align} 
	\alpha_D\equiv\varphi_{OS}+\iint_{\Omega_O+\Omega_S} K dS+\int_{C_0}\kappa_g d\ell. 
\end{align} 
Along $C_0$, $\frac{r_0}{\sqrt{\mathcal{A}(r_0)}} = b$ is a constant. Therefore, we have 
\begin{align}
	\begin{split}
		\int_{C_0} \kappa_g d\ell 
		&=-\left(\frac{a^2}{r_0}-\frac{3M }{r^2_0} \right)b\, \varphi_{OS}.
	\end{split}
	\label{eqn:kappaell}
\end{align}
Similar to the analysis in Appendix\ref{sec:appendix}, for the radial lines $\varphi = \varphi(r) = \text{const}$, we find 
\begin{align}
	\kappa_g=-\frac{1}{2\sqrt{\gamma_{\varphi\varphi}}}\frac{\partial \ln \gamma_{rr} }{\partial \varphi}=0.
\end{align}
The geodesic curvature along the light ray from source to observer, calculated using the generalized optical metric from Eq.~(\ref{eqn:opticalmetricA}), is also readily found to be zero.

As discussed in Sec.~\ref{sec:exact}, $\varphi_{OS}$ in the presence of the cosmological constant $\Lambda$ is identical to $\alpha_{1D}(b_{\mathrm{eff}})$; however, the light ray has a different turning point $r_0$, as shown in Eq.~(\ref{eqn:beff}). Therefore, from Eqs.~(\ref{eqn:drdphi}) and (\ref{eqn:dudphi}), it is clear that $\varphi_{OS}$ can be calculated using the same perturbative method as in the previous section, with the substitution $b \to b_{\mathrm{eff}}$: 
\begin{align}
	\begin{split}
		\varphi_{OS}&=\varphi_O-\varphi_S=\alpha_{1D}(b_{\mathrm{eff}}),
	\end{split}
	\label{eqn:varphiOS}
\end{align}
In other words, neglecting terms of order $\mathcal{O}\left(\frac{M^3}{b_{\mathrm{eff}}^3}\right)$ and higher, we have 
\begin{align}
	\begin{split}
		u(\varphi) &\simeq u^{(0)}(\varphi) + \frac{M}{b_{\mathrm{eff}}} u^{(1)}(\varphi) + \frac{M^2}{b_{\mathrm{eff}}^2} u^{(2)}(\varphi) + \frac{M^3}{b_{\mathrm{eff}}^3} u^{(3)}(\varphi)+ \frac{M^4}{b_{\mathrm{eff}}^4} u^{(4)}(\varphi) ,
	\end{split}
	\label{eqn:uphiB}
\end{align}
where $u^{(0)}(\varphi)$, $u^{(1)}(\varphi)$, and $u^{(2)}(\varphi)$ are given in Eqs.~(\ref{eqn:uphi}) and (\ref{eqn:uphi012}), with $b$ replaced by $b_{\mathrm{eff}}$.

The Gaussian curvature $K$ can be computed as 
\begin{align}
	\begin{split}
		K&=\frac{1}{\sqrt{\mathrm{det}(\gamma_{IJ})}} \left[{\partial_\varphi}\left(\frac{\partial_\varphi \sqrt{\gamma_{rr}}} {\sqrt{\gamma_{\varphi\varphi}}}\right)- {\partial _r}\left( \frac{\partial_r \sqrt{\gamma_{\varphi\varphi}}} {\sqrt{\gamma_{rr}}}   \right)\right].
	\end{split}
\end{align}
Substituting the new optical metric introduces an additional term compared to Eq.~(\ref{eqn:K}):
\begin{align}
	K=-\frac{2a^2 M}{r^3}+\frac{3 M^2}{r^4}-\left(\frac{a^2}{3}-\frac{2 M}{r}\right)\Lambda.
	\label{eqn:K2}
\end{align}
The area element on the equatorial plane is given by 
\begin{align}
	\begin{split}
		dS & =\sqrt{\mathrm{det}(\gamma_{IJ})}drd\varphi =\frac{r dr d\varphi}{\left(a^2 - \frac{2M}{r} - \frac{\Lambda r^2}{3}\right)^{{3}/{2}}}.
	\end{split}
\end{align}
Note that the integration over $r$ can actually be performed exactly in closed form:
\begin{align}
	\begin{split}
		\iint_{\Omega_O+\Omega_S} K dS &= - \int^{\varphi_O}_{\varphi_S} d\varphi \left. \frac{a^2 - \frac{3M}{r}}{\sqrt{a^2 - \frac{2M}{r} - \frac{\Lambda r^2}{3}}} \right|_{r_0}^{r(\varphi)}.
	\end{split}
\end{align}
However, this exact intermediate result is not particularly helpful for the integration over $\varphi$ in the weak-field limit. Therefore, it is convenient to apply the weak-field approximation at this stage:
\begin{align}
	\begin{split}
		dS	&\simeq \frac{r}{a^3} \left[1 + \frac{3M}{a^2 r} + \frac{15M^2}{2a^4 r^2}  + \frac{35M^3}{2a^6 r^3} + \frac{315M^4}{16a^8 r^4}\right. \\
		&\qquad\quad + \frac{\Lambda }{a^2}\left(\frac{r^2}{2} + \frac{5 M r}{2 a^2 }  + \frac{35 M^2 }{4 a^4 } \right)
		\left.  + \mathcal{O}\left(\frac{M^5}{r^5}, \frac{M^3\Lambda}{r},\Lambda^2 r^4\right)\right] dr d\varphi.
	\end{split}
	\label{eqn:dS}
\end{align}

Combining Eqs.~(\ref{eqn:K2}) and (\ref{eqn:dS}), we obtain 
\begin{align}
	\begin{split}
		\iint_{\Omega_O+\Omega_S} K dS 
		&\simeq T_1 - T_2 + \mathcal{O}\left(\frac{M^4}{r^4_0}, \frac{M^3 \Lambda}{r_0}, \Lambda^2 r_{O/S}^4  \right),
	\end{split}
	\label{eqn:KSTT}
\end{align}
where we define 
\begin{align}
	\begin{split}
		T_1 &= \int^{\varphi_O}_{\varphi_S} d\varphi \int^{r(\varphi)}_{r_0} dr \left(-\frac{2 M}{ar^2} - \frac{3 M^2}{a^3r^3} - \frac{6 M^3}{a^5 r^4} - \frac{25 M^4}{2a^7 r^5}\right) \\
		&=\int^{\varphi_O}_{\varphi_S} d\varphi \left.\left(\frac{2M}{a} u + \frac{3 M^2}{2a^3}u^2 + \frac{2M^3}{a^5}u^3+ \frac{25M^4}{8a^7}u^4 \right)\right|_{u_0}^{u(\varphi)}\\
		&=T_{1\varphi}-T_{10},
	\end{split}
\end{align}
and $T_2$ as the term proportional to $\Lambda$: 
\begin{align}
	\begin{split}
		T_2 &= \Lambda \int^{\varphi_O}_{\varphi_S} d\varphi \int^{r(\varphi)}_{r_0} \frac{r dr}{3a} \\
		&= \Lambda  \int^{\varphi_O}_{\varphi_S} d\varphi \left. \frac{1}{6au^2}\right|_{u_0}^{u(\varphi)}\\
		&=T_{2\varphi} - T_{20}.
	\end{split}
\end{align}

Plug in Eqs.~(\ref{eqn:uphiB}) and (\ref{eqn:varphiOS}) into the first part of $T_1$, we get
\begin{align}
	\begin{split}
		T_{1\varphi} & =\int^{{\varphi_O}}_{\varphi_S} d\varphi\left[ \frac{2M }{a}u(\varphi) + \frac{3M^2 }{2a^3}u^2(\varphi)  + \frac{2M^3}{a^5}u^3(\varphi)  + \frac{25M^4}{8a^7}u^4(\varphi) \right]   \\
		&\simeq \frac{4 M}{a^3 b_{\mathrm{eff}}}
		+ \frac{15\pi  M^2}{ 4 a^6 b_{\mathrm{eff}}^2} + \frac{128  M^3}{3 a^9 b_{\mathrm{eff}}^3} + \frac{3465 \pi M^4}{64 a^{12} b_{\mathrm{eff}}^4}  		- \frac{M b_{\mathrm{eff}}}{a }\left(\frac{1}{r^2_O}+\frac{1}{r^2_S}\right)	- \frac{M^2 b_{\mathrm{eff}}}{2a^3 }\left(\frac{1}{r^3_O}+\frac{1}{r^3_S}\right).
	\end{split}
\end{align}
The second part of $T_1$ is simply
\begin{align}
	\begin{split}
		T_{10} 
		&= \int^{{\varphi_O}}_{\varphi_S} d\varphi\left(\frac{2M}{a} u_0 + \frac{3M^2}{2a^3} u^2_0  + \frac{2M^3}{a^5}u^3_0  + \frac{25M^4}{8a^7}u^4_0\right)   \\
		&=\left(\frac{2M}{a r_0} + \frac{3M^2}{2a^3 r^2_0} + \frac{2M^3}{a^5 r_0^3} + \frac{25M^4}{8a^7r_0^4} \right) \varphi_{OS}.
	\end{split}
\end{align}

It is nontrivial to compute $T_{2\varphi}$,
\begin{align}
	\begin{split}
		T_{2\varphi} &=\frac{\Lambda}{6a}  \int^{\varphi_O}_{\varphi_S} \frac{d\varphi}{u^2(\varphi)}.
	\end{split}
\end{align}
One might naively attempt to expand the integrand in powers of $M/b$ and perform the integration term by term. However, this approach is incorrect because the leading approximation, $u^{(0)}(\varphi)$, appears in the denominator and may become small. The correct asymptotic expansion involves retaining $u^{(0)}(\varphi) + \frac{M}{b} u^{(1)}(\varphi)$ in the denominator, while expanding the remaining terms in powers of $M/b$. To fourth order, we have
\begin{align}
	\begin{split}
		\frac{1}{u^2(\varphi)} &\simeq \frac{1}{\left[u^{(0)}(\varphi) + \frac{M}{b} u^{(1)}(\varphi)\right]^2} \left\{ 1 -  \frac{2}{u^{(0)}(\varphi) + \frac{M}{b} u^{(1)}(\varphi)} \left[\frac{M^2}{b^2} u^{(2)}(\varphi)   + \frac{M^3}{b^3} u^{(3)}(\varphi) + \frac{M^4}{b^4} u^{(4)}(\varphi)\right] \right. \\
		&\quad \left. +  \frac{3}{\left[u^{(0)}(\varphi) + \frac{M}{b} u^{(1)}(\varphi)\right]^2} \left[\frac{M^2}{b^2} u^{(2)}(\varphi) \right]^2 \right\}
	\end{split}
\end{align}
Since $T_2$ is proportional to $\Lambda$, we compute $T_2$ to order $M^2\Lambda$. To achieve this, we first integrate over $\varphi$ in $T_{2\varphi}$, and then perform a \textit{multiple-scale asymptotic expansion} involving two sets of small parameters: $M/b$ and $b/r_{O/S}$ \cite{BenderBook}. According to Eqs.~(\ref{eqn:DphiS}) and (\ref{eqn:DphiO}), one possible consistent expansion is 
\begin{align}
	\begin{split}
		\frac{M}{b} \sim \epsilon, \quad \Delta\varphi_{O/S} \simeq \frac{b}{r_{O/S}} \sim \delta\sim  \sqrt{\epsilon}.  
	\end{split}
\end{align}
To orders $M^2\Lambda$, we obtain
\begin{align}
	\begin{split}
		T_{2\varphi} &\simeq \frac{\Lambda}{6a} \left[b(r_O+r_S) - \frac{a^2b^3}{2} \left(\frac{1}{r_O} + \frac{1}{r_S}\right)  - \frac{a^4b^5}{8} \left(\frac{1}{r_O^3} + \frac{1}{r_S^3}\right)  + \frac{2Mb}{a^2}  + \frac{Mb^3}{2} \left(\frac{1}{r_O^2} + \frac{1}{r_S^2}\right)
		  + \frac{3\pi M^2}{2 a^5}\right].
	\end{split}
\end{align}
By applying the perturbative solution for $u_0$ from Eq.~(\ref{eqn:utildeb}), the second part of $T_2$ is given by
\begin{align}
	T_{20} &=\frac{\Lambda}{6au_0^2}  \int^{\varphi_O}_{\varphi_S} d\varphi 
\simeq  \frac{\Lambda a b^2}{6} \left( 1 - \frac{2M}{a^3b}  - \frac{2M^2}{a^6b^2} \right)\varphi_{OS}.
\end{align}

Put all the pieces in Eqs.~(\ref{eqn:kappaell}), (\ref{eqn:varphiOS}), and (\ref{eqn:KSTT}) together, and we obtain the deflection angle in the weak-field limit.  
Plug $T_1$ into Eq.~(\ref{eqn:KSTT}) and combine it with Eq.~(\ref{eqn:kappaell}), and we have
\begin{align}
	\begin{split}
		\alpha_D(b,\Lambda) 
		&\simeq
		(1-a)\frac{\pi}{a} + \frac{4 M}{a^4 b} 
		+ \frac{ 15 \pi M^2}{4a^7 b^2} + \frac{ 128 M^3}{3a^{10} b^3}  + \frac{ 3465  \pi M^4}{64a^{13} b^4}
		-(1-a) b \left(\frac{1}{r_O}+\frac{1}{r_S}\right)  - \frac{(1-a)a^2b^3}{6} \left(\frac{1}{r_O^3}+\frac{1}{r_S^3}\right) \\
		&\quad
		- \frac{Mb}{a} \left(\frac{1}{r_O^2}+\frac{1}{r_S^2}\right)
		- \frac{\Lambda b (r_O+r_S)}{6a}  + \left(2 - a\right)\frac{\Lambda b^3 }{12} \left(\frac{1}{r_O}+\frac{1}{r_S}\right) + (4-3a)a^2\frac{ \Lambda b^5 }{48} \left(\frac{1}{r_O^3}+\frac{1}{r_S^3}\right) 
		\\
		&\quad  + (2-a)\frac{M \Lambda b}{3 a^4 }  + \frac{M \Lambda b^3 }{12a} \left(\frac{1}{r_O^2}+\frac{1}{r_S^2}\right)  + \frac{5\pi M^2 \Lambda}{4 a^7 }  - \frac{M^2 \Lambda b }{4a^5} \left(\frac{1}{r_O}+\frac{1}{r_S}\right) \\
		&\quad 
		+ \mathcal{O}\left(   \frac{M^5}{b^5}, \frac{M^4}{b^3r_{O/S}}, \frac{M^3}{br_{O/S}^2},\frac{M^2b}{r_{O/S}^3},\frac{Mb^3}{r_{O/S}^4},\frac{b^5}{r_{O/S}^5}, \frac{M^3 \Lambda}{b}, \Lambda^2 b r_{O/S}^3 \right).
	\end{split}
\end{align}
which matches the result in Eq.~(\ref{eqn:alphaDweak}) term by term.

\end{document}